\documentclass[12pt,a4paper]{article}
\usepackage[margin=1in]{geometry}
\usepackage{setspace}
\usepackage[utf8]{inputenc}
\usepackage{tikz} 

\usepackage{multicol}
\usepackage{caption} 
\usepackage{subcaption}
\usepackage[inline]{enumitem}

\usepackage{amsmath, amssymb, mathtools, amsthm, bbm}

\newtheorem{theorem}{Theorem}[section]
\newtheorem{proposition}[theorem]{Proposition}

\newtheorem{corollary}[theorem]{Corollary}

\usepackage{graphicx}
\usepackage{tikz}
\usepackage{circuitikz}
\usepackage{pgf}
\usetikzlibrary{
  arrows, shapes.arrows, shapes.geometric,
  shapes.multipart, decorations.pathmorphing,
  positioning, swigs
}

\usepackage[dvipsnames]{xcolor}

\usepackage[normalem]{ulem}  
\usepackage{algpseudocode}
\usepackage{url}
\usepackage{booktabs}
\usepackage{hyperref}
\usepackage{comment}
\usepackage{ytableau}
\ytableausetup{centertableaux}
\usepackage{tkzexample}
\usepackage{enumitem}

\usepackage[style=apa, backend=biber, natbib=true, maxcitenames=3, mincitenames=1, maxbibnames=6, minbibnames=6, uniquename=false, uniquelist=false]{biblatex}
\usepackage{xcolor}

\usepackage{glossaries}
\newacronym{rct}{RCT}{randomised controlled trial}
\newacronym{gp}{GP}{Gaussian process}
\newacronym{hte}{HTE}{heterogeneous treatment effect}
\newacronym{ate}{ATE}{Average Treatment Effect}
\newacronym{cate}{CATE}{conditional average treatment effect}
\newacronym{icm}{ICM}{Intrinsic Coregionalization Model}
\newacronym{rwd}{RWD}{Real-World Data}
\newacronym{cf}{CF}{Causal Forest}
\newacronym{rf}{RF}{Random Forest}
\newacronym{rmse}{RMSE}{Root Mean Squared Error}
\newacronym{rlm}{RLM}{Random Location Model}
\newacronym{osm}{OSM}{Outcome Sensitivity Model}

\newcommand{\papertitle}{Bayesian Nonparametric Sensitivity Analysis for Unmeasured Confounding in Data Fusion}
\title{\papertitle}
\author{Evangelos Dimitriou}

\usepackage{fancyhdr}
\begin{document}

\begin{center}
    {\Large\bfseries \papertitle\par}
    \vspace{1.5em}

    {\large Evangelos Dimitriou$^{1}$,
    Edwin Fong$^{2}$,
    Jens Magelund Tarp$^{3}$,
    Karla Diaz-Ordaz$^{1}$,
    Brieuc Lehmann$^{1}$}\\[0.75em]

    {\small
    $^{1}$Department of Statistical Science, University College London\\
    $^{2}$Department of Statistics and Actuarial Science, University of Hong Kong\\
    $^{3}$Novo Nordisk A/S}
\end{center}

\vspace{1em}

\begin{singlespace}
\noindent\textbf{Corresponding author:}\\
Evangelos Dimitriou\\
Email: \href{mailto:evangelos.dimitriou.22@ucl.ac.uk}{evangelos.dimitriou.22@ucl.ac.uk}
\end{singlespace}

\vspace{0.5em}

\begin{center}\textbf{Abstract}\end{center}
\vspace{-1em}
\begingroup
\let\newpage\relax
\let\clearpage\relax
\let\pagebreak\relax
\let\cleardoublepage\relax

Causal conclusions from observational data rest on assumptions, such as exchangeability over treatment, that cannot be verified from the data alone. 
Sensitivity analysis asks how conclusions change as such assumptions are relaxed. 
We develop a Bayesian nonparametric sensitivity analysis framework based on \emph{Causal-ICM}, a multi-task Gaussian process model that places a prior on the causal response surface and the observational outcome regression. 
Their prior correlation is the key sensitivity parameter, where each value returns a full posterior for the average treatment effect, rather than only a bound or a point estimate. 
The framework applies to observational data alone 
and extends directly to settings where a randomised trial is available alongside a larger, potentially confounded observational study.
Full pooling and the trial-only analysis then emerge as endpoints of the same sensitivity analysis. 
The posterior mean of the average treatment effect decomposes into a trial-only component plus a shift due to observational data, and for any correlation below one, posterior uncertainty remains bounded away from zero however large the observational study becomes. 
We illustrate the framework with simulations and a synthetic-data example combining a randomised trial of oral semaglutide with real-world health data, in which the trial's conclusion holds across the entire sensitivity range, regardless of how much the observational data are trusted.

\newpage

\endgroup

\newpage

\setlength{\parindent}{20pt}

\section{Introduction}\label{sec:introduction}

Assessing the robustness of causal conclusions to violations of identifying assumptions, such as unmeasured confounding, is a central task in causal inference from observational data. 
Causal sensitivity analysis addresses this task by examining how conclusions change as such assumptions are relaxed \citep{cornfield1959smoking, Rosenbaum2002}.
Most existing frameworks fit a fixed sensitivity model to a single study and report either a worst-case bound or a point estimate for each hypothesised degree of violation, for example bounds on treatment-assignment bias \citep{tan2005MSM, zhao2018sensitivityanalysisinverseprobability}, bounds on outcome-model discrepancies \citep{bonvini2022sensitivityanalysismarginalstructural,franks2019flexiblesensitivityanalysisobservational, zhang2025enhancedmarginalsensitivitymodel} and potential outcome distributions \citep{nabi2024}, or interpretable scalar summaries such as the E-value \citep{vanderweeleEvalue} and the robustness value \citep{cinelli2019}. 


A Bayesian causal sensitivity analysis treats the magnitude of an assumption violation, such as confounding bias, as an unidentified parameter and assigns it a prior \citep{mccandless2007bayesian, dhawan2026bayesian}. This differs from Bayesian prior sensitivity analysis in identified models \citep{berger1994overview} where the influence of the prior typically diminishes as the sample grows. In a causal setting, however, assumption violations may remain only partially identified, so prior assumptions can continue to influence causal conclusions even in large samples \citep{gustafson2010bayesian}. The posterior for the causal effect therefore combines sampling variability, which shrinks with sample size, and uncertainty about the bias, which does not. Existing approaches either rely on parametric outcome models \citep{mccandless2007bayesian}, or place a prior only on the violation and use it to assess the sensitivity of a frequentist estimate, without updating the prior with the study data \citep{dhawan2026bayesian}. Others fit flexible Bayesian outcome models but fix the sensitivity parameters and report a posterior at each value \citep{dorie2016flexible, franks2019flexiblesensitivityanalysisobservational}.

The growing availability of large-scale observational studies has sparked interest in data fusion techniques that combine potentially confounded real-world data with randomised experiments. This body of literature addresses the fundamental tension between \glspl{rct}, which provide internal validity but are often small and may not be representative of the target population, and observational studies that may be more reflective of ongoing clinical practice but may be affected by unmeasured confounding \citep{Stuart2011, hernan2020causal}. Several approaches for point estimation have been proposed to manage unmeasured confounding in the observational study: testing compatibility before pooling \citep{yang2022elasticintegrativeanalysisrandomized, yang2026datafusiondistributionalequivalence}, tempering the influence of the observational data on the analysis \citep{lin2024combiningexperimentalobservationaldata, dimitriou2025causalicmdatafusionframework}, using the trial to correct observational estimates \citep{kallus2018removinghiddenconfoundingexperimental, yang2022integrative}, or using the trial to detect bias in the observational study \citep{fawkes2025hardnessvalidatingobservationalstudies}; see \citet{colnet2023causalinferencemethodscombining} and \citet{lin2024datafusionefficiencygain} for reviews. Moving beyond point identification, recent frameworks derive partial-identification bounds under explicit violation parameters \citep{lanners2025datafusionpartialidentification} or use external data that measure the unmeasured confounder to undertake probabilistic sensitivity analysis \citep{comment2022bayesian}.

This paper develops a Bayesian sensitivity analysis framework for the \gls{ate} that relaxes the assumption of no unmeasured confounding in the observational study, so that the observational outcome regression is informative about the causal response surface identified by the randomised trial. Building on Gaussian process approaches to causal inference and heterogeneous treatment effect estimation \citep{alaa2017bayesianinferenceindividualizedtreatment, witty2020causalinferenceusinggaussian}, we jointly model the trial and observational outcome regressions under the \emph{Causal-ICM} framework introduced in \citet{dimitriou2025causalicmdatafusionframework}. 
When only observational data are available, our approach is a Bayesian analogue to the \gls{osm} of \citet{bonvini2022sensitivityanalysismarginalstructural} in its focus on departures of the observational outcome regressions from the causal ones. Where the \gls{osm} bounds these departures, Causal-ICM induces a prior on them and so yields a posterior for the causal effect; it also extends naturally to the data fusion setting when trial data are also available. It builds on work that places priors over assumption violations \citep{dhawan2026bayesian} and on methods that borrow from observational data in data fusion \citep{lin2024combiningexperimentalobservationaldata}.
We encode sensitivity through a prior correlation, yielding closed-form posteriors across the full sensitivity spectrum.
We show that: (1) under \emph{Causal-ICM}, the posterior mean and variance of the causal effect decompose into an \gls{rct}-only term and an observational contribution; (2) the shift in the posterior mean induced by observational data does not grow with the observational sample size; and (3) when trial data are available, the confounding bias of the observational study, which is unidentifiable in the single-study setting, is identifiable and has a closed-form posterior under \emph{Causal-ICM}.

Section~\ref{sec:problem_setup_assumptions} sets out notation and identifying assumptions; Section~\ref{sec:preliminaries} reviews multi-task Gaussian processes and Section~\ref{sec:model} presents \emph{Causal-ICM}, and Sections~\ref{sec:obs_only}--\ref{sec:rct-data} characterise its posterior without and with trial data, establishing identifiability and bounds on the influence of the observational data; Section~\ref{sec:bias_identification} shows that the trial identifies the confounding bias; Sections~\ref{sec:simulation-study}--\ref{sec:pioneer6-application} illustrate the framework in simulations and in synthetic data based on the PIONEER~6 clinical trial; Section~\ref{sec:discussion} concludes with a discussion.
\section{Notation, Assumptions, and Identification}\label{sec:problem_setup_assumptions}

\begin{figure}[htbp]
    \setstretch{1}
    \centering
    \begin{subfigure}[b]{0.45\textwidth}
        \centering
        \begin{tikzpicture}[
            node distance=1.2cm and 1.5cm,
            every node/.style={circle, draw=black, thick, minimum size=7mm, font=\small},
            dashedNode/.style={dashed}, 
            arrow/.style={-Stealth, thick}
        ]
            \node (X) {$X$};
            \node (A) [below left=of X] {$A$};
            \node (Y) [below right=of X] {$Y$};
            \node (U) [below right=of A, dashedNode] {$U$}; 
            
            \draw [arrow] (A) -- (Y);
            \draw [arrow] (X) -- (Y);
            \draw [arrow] (X) -- (A);
            \draw [arrow] (U) -- (Y);
        \end{tikzpicture}
        \caption{RCT}
        \label{fig:dag_left}
    \end{subfigure}
    \hfill 
    \begin{subfigure}[b]{0.45\textwidth}
        \centering
        \begin{tikzpicture}[
            node distance=1.2cm and 1.5cm,
            every node/.style={circle, draw=black, thick, minimum size=7mm, font=\small},
            dashedNode/.style={dashed}, 
            arrow/.style={-Stealth, thick}
        ]
            \node (X) {$X$};
            \node (A) [below left=of X] {$A$};
            \node (Y) [below right=of X] {$Y$};
            \node (U) [below right=of A, dashedNode] {$U$};
            
            \draw [arrow] (A) -- (Y);
            \draw [arrow] (X) -- (Y);
            \draw [arrow] (X) -- (A);
            
            \draw [arrow] (U) -- (Y);
            \draw [arrow] (U) -- (A); 
        \end{tikzpicture}
        \caption{Observational Study}
        \label{fig:dag_right}
    \end{subfigure}

    \caption{Directed Acyclic Graphs representing the causal structure of (a) the RCT and (b) the observational study. The edge from \(X\) to \(A\) in the \gls{rct} represents stratified randomisation and it is included for completeness; under simple randomisation, this edge would be omitted.}
    \label{fig:dags}
\end{figure}
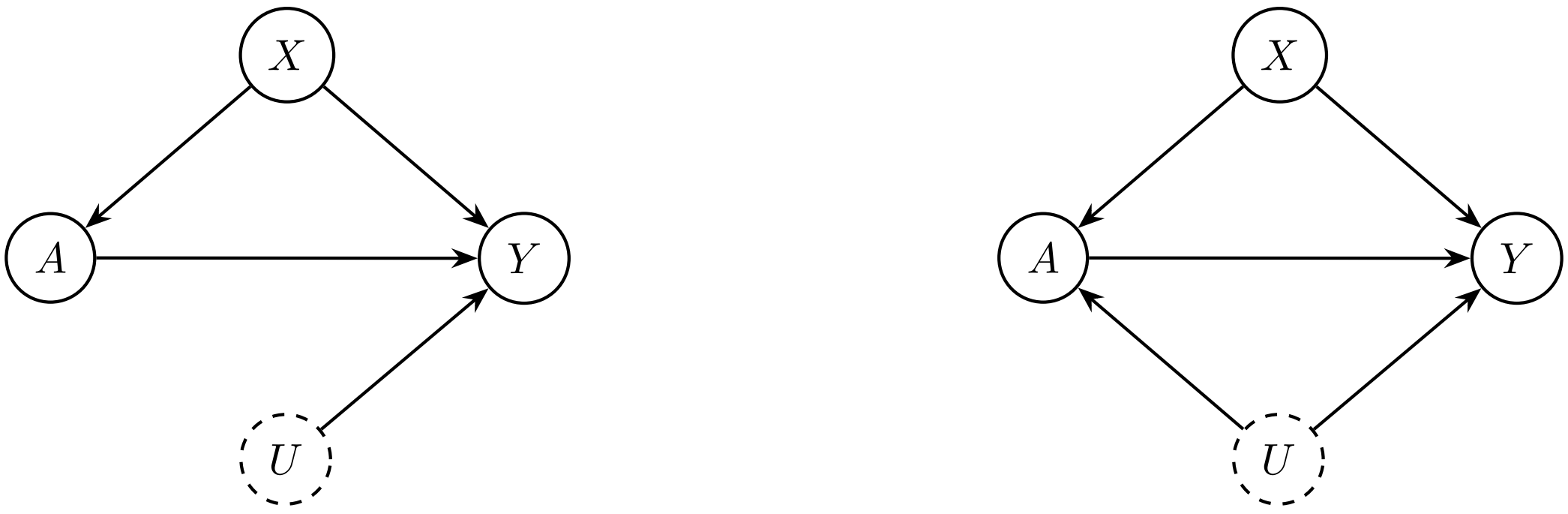

We consider two data sources: an \gls{rct}, denoted by $\mathcal{D}_{\mathrm{e}}$ of size $n_{\mathrm{e}}$, and an observational study, denoted by $\mathcal{D}_{\mathrm{o}}$ of size $n_{\mathrm{o}}$. The study indicator is $S \in \{\mathrm{e}, \mathrm{o}\}$, binary treatment is $A \in \{0,1\}$, baseline covariates are $X \in \mathcal{X}$, potential outcomes are $Y(a)$, and observed continuous outcomes \(Y \in \mathbb{R}\).
Let \(U\) be a (set of) latent variable(s) such that \((X,U)\) would be sufficient to control for confounding of the effect of \(A\) on \(Y\) were \(U\) observable in \(S=\mathrm{o}\).
Figure~\ref{fig:dags} represents the causal structure of our setting.

For \(s\in \{\mathrm{e},\mathrm{o}\}\) and for treatment level $a\in \{0,1\}$, let \(\nu_s(x,a) = \mathbb{E}[Y(a)\mid X=x,S=s]\) be the study-specific causal response surface, so that the study-specific \gls{cate} is $\tau_s(x) := \mathbb{E}[Y(1)-Y(0)\mid X=x,S=s] = \nu_s(x,1)-\nu_s(x,0)$. Let $\mu_{\mathrm{e}}(x,a) = \mathbb{E}[Y\mid X=x,A=a, S=\mathrm{e}]$ denote the experimental outcome response surface, and let $\mu_{\mathrm{o}}(x,a) = \mathbb{E}[Y\mid X=x,A=a, S=\mathrm{o}]$ denote the observational outcome regression function, for treatment arm $a\in\{0,1\}$.  For the observational study, define the latent $U$-conditional causal response surface $\mu_{\mathrm{o}}(x,u,a) = \mathbb{E}[Y(a)\mid X=x,U=u,S=\mathrm{o}]$. 

Identification assumptions connect these causal quantities to observable conditional means. The stable unit treatment value assumption (SUTVA) is assumed in both studies; the remaining assumptions are stated separately for the trial and observational data.

\begin{enumerate}[label=\textbf{A\arabic*}, itemjoin={{\, }}]
    \item \label{A1:consistency} \emph{SUTVA}: an individual's potential outcome depends only on their own treatment (no interference), treatment versions do not affect potential outcomes (no relevant hidden versions), and $A=a \Rightarrow Y=Y(a)$ for each $a\in\{0,1\}$ (consistency).
\end{enumerate}

Identifying $\mathbb{E}[Y(a)\mid X=x, S=\mathrm{e}]$ from the experimental data requires the following assumptions:
\begin{enumerate}[label=\textbf{A\arabic*}, itemjoin={{\, }}, resume] 
    \item \label{A2:RCT_A_exchangeability} \emph{Mean conditional exchangeability over $A$ given $X$ in the \gls{rct}}: for each $a\in\{0,1\}$, $\mathbb{E}[Y(a)\mid X=x,A=a,S=\mathrm{e}] = \mathbb{E}[Y(a)\mid X=x,S=\mathrm{e}]$, i.e.\ treatment assignment in the \gls{rct} is mean-independent of the potential outcomes given $X$, 
    \item \label{A3:RCT_A_positivity} \emph{Positivity in the \gls{rct}}: for each $a\in\{0,1\}$, $0 < \mathbb{P}(A=a\mid X=x, S=\mathrm{e}) <1$ for all $x$ in the support of $X\mid S=\mathrm{e}$.
\end{enumerate}

Under \ref{A1:consistency}--\ref{A3:RCT_A_positivity}, it follows that $\nu_{\mathrm{e}}(x,a) = \mu_{\mathrm{e}}(x,a)$, so that the CATE in the RCT can be identified with the difference in the outcome response surfaces, which are estimable from data:
\[
\tau_{\mathrm{e}}(x) = \mu_{\mathrm{e}}(x,1) - \mu_{\mathrm{e}}(x,0).
\]


Identification when using observational data requires additional assumptions
(Figure~\ref{fig:dags}). To link the two studies and use the observational data we make the following assumptions:
\begin{enumerate}[label=\textbf{A\arabic*}, itemjoin={{\, }}, resume]
    \item \label{A4:OBS_A_exchangeability} \emph{Latent mean conditional exchangeability over $A$ in the observational study}: for each $a\in\{0,1\}$, \(\mathbb{E}[Y(a)\mid X=x,U=u,S=\mathrm{o}] = \mathbb{E}[Y(a)\mid X=x,U=u,A=a,S=\mathrm{o}]\); we do not assume this holds conditional on $X$ alone, 
    \item \label{A5:OBS_A_positivity} \emph{Positivity in the observational study}: for each $a\in\{0,1\}$, $0<\mathbb{P}(A=a\mid X,U,S=\mathrm{o})<1$ almost surely, 
    \item \label{A5:S_exchangeability} \emph{Mean conditional exchangeability over $S$}: for each $a\in\{0,1\}$, $\mathbb{E}[Y(a)\mid X=x, S=\mathrm{e}] = \mathbb{E}[Y(a)\mid X=x, S=\mathrm{o}]$; this is marginal over $U$, requiring equality of the marginal causal response surfaces across studies given $X$, but not equality of the $U$-conditional surfaces, 
    \item \label{A6:S_positivity} \emph{Positivity of \gls{rct} participation}: $0 < \mathbb{P}(S=\mathrm{e}\mid X=x) < 1$ for all $x$ in the support of $X$.
\end{enumerate}

Assumption~\ref{A4:OBS_A_exchangeability} says that the combined covariate set \(\{X,U\}\) together suffice to control confounding in the observational study, though note in practice we only observe $X$. 
Assumption \ref{A5:S_exchangeability} implies $\nu_{\mathrm{e}}(x,a)=\nu_{\mathrm{o}}(x,a)$, and hence \(\tau_{\mathrm{e}}(x) = \tau_{\mathrm{o}}(x)\) on the common covariate support of the two studies. This assumption may fail, for example due to effects differing between populations or due to trial engagement effects \citep{ung2025generalizing, Dahabreh2019}. Combined with the \gls{rct} identification result, it gives $\nu_{\mathrm{o}}(x,a)=\mu_{\mathrm{e}}(x,a)$ in that common support, provided Assumption~\ref{A3:RCT_A_positivity} holds: the \gls{rct} identifies the observational-population causal response surface, without requiring $\mu_{\mathrm{o}}(x,a)$ itself to be causal. Indeed, using \ref{A1:consistency}, the law of total expectation, and \ref{A4:OBS_A_exchangeability},
\[
\mu_{\mathrm{o}}(x,a) = \int \mu_{\mathrm{o}}(x,u,a)\, dP(u\mid X=x,A=a,S=\mathrm{o}),
\]
whereas $\nu_{\mathrm{o}}(x,a) = \int \mu_{\mathrm{o}}(x,u,a)\, dP(u\mid X=x,S=\mathrm{o})$. In general $\mu_{\mathrm{o}}(x,a)\neq \nu_{\mathrm{o}}(x,a)$, because the observational regression averages the latent causal response over the treatment-arm-specific distribution of $U$, whereas the causal response surface averages it over the marginal distribution of $U$ given $X$ alone.

We define the arm-specific observational confounding bias as the resulting discrepancy,
\begin{equation*}
    \begin{aligned}
        d(x,a) &= \mu_{\mathrm{o}}(x,a)-\nu_{\mathrm{o}}(x,a) \\
        &= \int \mu_{\mathrm{o}}(x,u,a)\, dP(u\mid X=x,A=a,S=\mathrm{o}) - \int \mu_{\mathrm{o}}(x,u,a)\, dP(u\mid X=x,S=\mathrm{o}),
    \end{aligned}
\end{equation*}
which, by Assumption \ref{A5:S_exchangeability}, is equivalently $d(x,a) = \mu_{\mathrm{o}}(x,a)-\mu_{\mathrm{e}}(x,a)$: the discrepancy between the observational and the experimental response surfaces identifies the arm-specific confounding bias. Under \ref{A1:consistency}--\ref{A3:RCT_A_positivity} and \ref{A5:S_exchangeability}, and on the common support where both treatments have positivity probability in each study, $d(x,a)=0$ for $a\in\{0,1\}$ is equivalent to mean conditional exchangeability over treatment in the observational study given $X$.

Define the observational treatment contrast $\omega_{\mathrm{o}}(x) = \mu_{\mathrm{o}}(x,1)-\mu_{\mathrm{o}}(x,0)$. Since $\tau_{\mathrm{e}}(x) = \mu_{\mathrm{e}}(x,1)-\mu_{\mathrm{e}}(x,0)$, the bias decomposition above gives
\[
\omega_{\mathrm{o}}(x)-\tau_{\mathrm{e}}(x) = \{\mu_{\mathrm{o}}(x,1)-\mu_{\mathrm{e}}(x,1)\} - \{\mu_{\mathrm{o}}(x,0)-\mu_{\mathrm{e}}(x,0)\} = d(x,1)-d(x,0),
\]
so the discrepancy between the observational and experimental treatment contrasts is exactly the difference between the arm-specific confounding biases, which we henceforth denote  $D(x) = d(x,1)-d(x,0)$.

The \gls{cate} provides the building block for the population summaries considered below. For a target distribution \(P\), the \gls{ate} is \(\tau_P = \int \tau_{\mathrm{e}}(x) \; dP(x)\). Likewise, for a baseline-defined subgroup (G), the subgroup average treatment effect is \(\tau_G=\int \tau_{\mathrm{e}}(x)\,dP(x\mid X\in G)\). Both estimands average the same function $\tau_{\mathrm{e}}$ and differ only in the weighting over different covariate distributions.


The discrepancy $d(x,a)$ quantifies how much the observational regression differs from the causal response surface identified by the trial. Pooling the studies amounts to assuming $d \equiv 0$, while a trial-only analysis discards the observational data. Between these extremes, we now seek a joint model for $\mu_{\mathrm{e}}$ and $\mu_{\mathrm{o}}$ that allows $d$ to be non-zero and specifies how closely the two surfaces are related a priori. The next section introduces such a model, \emph{Causal-ICM}, in which a sensitivity parameter sets the prior correlation between the two surfaces.

\section{Methodology}
\label{sec:methodology}

\emph{Causal-ICM} models the experimental and observational outcome regressions jointly as correlated Gaussian processes. Their prior correlation, \(\rho\), is the sensitivity parameter that controls how strongly observational data may inform the causal surface identified by the \gls{rct}. We first review multi-task Gaussian processes, on which Causal-ICM is built.

\subsection{Preliminaries: Multi-task Gaussian Processes}
\label{sec:preliminaries}


A \gls{gp} is a collection of random variables such that any finite subset of the variables has a joint Gaussian distribution. It is completely specified by a mean function and a covariance function; we take the mean to be zero throughout. For brevity we drop the arm index below, writing $\mu_{\mathrm{e}}(x)$ and $\mu_{\mathrm{o}}(x)$ unless stated otherwise. 

For each treatment arm,
\[
\boldsymbol \mu = \Big( \mu_{\mathrm{e}}(x), \mu_{\mathrm{o}}(x) \Big)^\top \sim \mathcal{GP} \Big( \mathbf{0}, K \Big), 
\]
where \(K:\mathcal{X} \times \mathcal{X} \rightarrow \mathbb{R}^{2\times 2}\) is a matrix-valued covariance kernel whose \((s,s')\) entry is 
\[
K_{ss'}(x,x') = \operatorname{Cov}\{\mu_s(x), \mu_{s'}(x')\}, \quad s,s' \in \{\mathrm{e},\mathrm{o}\}.
\]
Thus, for any inputs \(x_1, \ldots, x_n \in \mathcal X\), where \(n=n_{\mathrm{e}}+n_{\mathrm{o}}\), and \(n_{\mathrm{e}}\) and \(n_{\mathrm{o}}\) the experimental and observational sample sizes, the stacked vector
\[\bigl(\boldsymbol{\mu}(x_1)^\top,\ldots,\boldsymbol{\mu}(x_n)^\top\bigr)^\top
\]
is a multivariate Gaussian with block covariance matrix whose \((i,j)\) block is \(K(x_i,x_j)\). The off-diagonal entries \(K_{\mathrm{eo}}\) and \(K_{\mathrm{oe}}\) encode prior dependence between the experimental and observational regressions, and hence determine the extent of information sharing across the two studies.


As usual, we assume a Gaussian observation model,
\begin{equation*}
    \begin{aligned}
        y_{\mathrm{e},i} = \mu_{\mathrm{e}}(x_i) + \varepsilon_{\mathrm{e},i}, \qquad
        y_{\mathrm{o},j} = \mu_{\mathrm{o}}(x_j) + \varepsilon_{\mathrm{o},j},
    \end{aligned}
\end{equation*}
with $\varepsilon_{\mathrm{e},i}, \varepsilon_{\mathrm{o},j} \sim \mathcal{N}(0,\sigma^2)$ independently across tasks and observations, $i=1,\dots,n_{\mathrm{e}}$, $j=1,\dots,n_{\mathrm{o}}$; a common noise variance across studies is assumed for simplicity but is easily relaxed.

The \gls{icm} is a special case of a multi-task \gls{gp} with a separable kernel, in which each task function is a linear combination of shared independent latent functions,
\begin{equation*}
    \begin{aligned}
        \mu_{\mathrm{e}}(x) = \sum_{q=1}^Q \alpha_{\mathrm{e},q} u_{q}(x), \quad
        \mu_{\mathrm{o}}(x) = \sum_{q=1}^Q \alpha_{\mathrm{o},q} u_{q}(x),
    \end{aligned}
\end{equation*}
where $u_1,\dots,u_Q \sim \mathcal{GP}(0,k)$ are i.i.d.\ for a chosen kernel $k$, and we set $Q=2$. The resulting multi-task covariance function factorises as
\begin{equation*}
    K(x,x')=B \otimes k(x,x'), \qquad
    B_{ss'}=\sum_{q=1}^Q\alpha_{s,q}\alpha_{s',q},
\end{equation*}
where $B$ is the coregionalisation matrix, encoding the prior covariance between tasks.


\subsection{\emph{Causal-ICM}: a \(\rho\)-Coupled Joint Prior}
\label{sec:model}

We define \emph{Causal-ICM} by fixing these coefficients as
\[
\begin{bmatrix}
\alpha_{\mathrm{e},1} & \alpha_{\mathrm{e},2} \\
\alpha_{\mathrm{o},1} & \alpha_{\mathrm{o},2}
\end{bmatrix} = \begin{bmatrix}
1 & 0 \\
\rho & \sqrt{1-\rho^2}
\end{bmatrix}, \qquad \rho \in [0,1],
\]
so that
\begin{equation*}
    \begin{aligned}
        \mu_{\mathrm{e}}(x) = u_1(x), \quad
        \mu_{\mathrm{o}}(x) = \rho\, u_1(x) + \sqrt{1-\rho^2}\, u_2(x).
    \end{aligned}
\end{equation*}
Under this parameterisation the coregionalisation matrix coincides with the correlation matrix,
\[
B = \begin{bmatrix}
1 & \rho \\
\rho & 1
\end{bmatrix}.
\]
Thus, \(\rho\) is the pointwise prior correlation between the experimental and observational outcome functions at every \(x\), controlling information sharing between the two studies. Larger values favour closer agreement between the surfaces, whereas smaller values allow greater departures.

Having specified the \(\rho\)-coupled prior, we now characterise the posterior it induces, before leveraging that posterior for sensitivity analysis with only observational data (Section~\ref{sec:obs_only}) and data fusion (Section~\ref{sec:rct-data}). Let \(\mathcal D=(y_{\mathrm{e}},X_{\mathrm{e}},y_{\mathrm{o}},X_{\mathrm{o}})=\{\mathcal D_{\mathrm{e}},\mathcal D_{\mathrm{o}}\}\) denote the experimental and observational data, and let \(x_*\) be a test point. The following proposition gives the joint posterior distribution of the experimental and observational response surfaces, in closed form and for any \(\rho\in[0,1]\). 

\begin{proposition}[Joint posterior distribution of \(\boldsymbol \mu(x_*)\), \textit{Causal-ICM}]
\label{prop:joint_post_distr}
Let
\[
\Sigma_{\mathrm{e}}=k(X_{\mathrm{e}},X_{\mathrm{e}})+\sigma^2I,\qquad
\Sigma_{\mathrm{eo}}=k(X_{\mathrm{o}},X_{\mathrm{o}})-\rho^2k(X_{\mathrm{o}},X_{\mathrm{e}})\Sigma_{\mathrm{e}}^{-1}k(X_{\mathrm{e}},X_{\mathrm{o}})+\sigma^2I,
\]
\[
r_{\mathrm{o}}=y_{\mathrm{o}}-\rho\,k(X_{\mathrm{o}},X_{\mathrm{e}})\Sigma_{\mathrm{e}}^{-1}y_{\mathrm{e}},\qquad
k'(x_*,x)=k(x_*,X_{\mathrm{e}})\Sigma_{\mathrm{e}}^{-1}k(X_{\mathrm{e}},x),
\]
\[
q(x_*)=k(x_*,X_{\mathrm{o}})-k'(x_*,X_{\mathrm{o}}),\qquad
a_{\mathrm{o}}(x_*)=k(x_*,X_{\mathrm{o}})-\rho^2k'(x_*,X_{\mathrm{o}}).
\]
Under \emph{Causal-ICM}, the joint posterior distribution of \(\boldsymbol \mu(x_*)=(\mu_{\mathrm{e}}(x_*),\mu_{\mathrm{o}}(x_*))^\top\) is bivariate normal,
\[
\boldsymbol \mu(x_*)\mid\mathcal D\sim\mathcal N\!\left(\begin{pmatrix}m_{\mathrm{e}|\mathcal D}(x_*)\\ m_{\mathrm{o}|\mathcal D}(x_*)\end{pmatrix},\;\begin{pmatrix}V_{\mathrm{e}|\mathcal D}(x_*)&C_{\mathrm{eo}|\mathcal D}(x_*)\\ C_{\mathrm{eo}| \mathcal D}(x_*)&V_{\mathrm{o}| \mathcal D}(x_*)\end{pmatrix}\right),
\]
with
\begin{align*}
m_{\mathrm{e}|\mathcal D}(x_*)&=k(x_*,X_{\mathrm{e}})\Sigma_{\mathrm{e}}^{-1}y_{\mathrm{e}}+\rho\,q(x_*)\Sigma_{\mathrm{eo}}^{-1}r_{\mathrm{o}},\\
m_{\mathrm{o}|\mathcal D}(x_*)&=\rho\,k(x_*,X_{\mathrm{e}})\Sigma_{\mathrm{e}}^{-1}y_{\mathrm{e}}+a_{\mathrm{o}}(x_*)\Sigma_{\mathrm{eo}}^{-1}r_{\mathrm{o}},\\
V_{\mathrm{e}|\mathcal D}(x_*)&=k(x_*,x_*)-k'(x_*,x_*)-\rho^2\,q(x_*)\Sigma_{\mathrm{eo}}^{-1}q(x_*)^\top,\\
V_{\mathrm{o}|\mathcal D}(x_*)&=k(x_*,x_*)-\rho^2k'(x_*,x_*)-a_{\mathrm{o}}(x_*)\Sigma_{\mathrm{eo}}^{-1}a_{\mathrm{o}}(x_*)^\top,\\
C_{\mathrm{eo}| \mathcal D}(x_*)&=\rho\{k(x_*,x_*)-k'(x_*,x_*)\}-\rho\,q(x_*)\Sigma_{\mathrm{eo}}^{-1}a_{\mathrm{o}}(x_*)^\top .
\end{align*}
\end{proposition}

\begin{proof}
The proof is provided in Web Appendix~\ref{app:proof:joint_post_distr}.
\end{proof}

Once the causal quantities have been identified under the assumptions stated above, inference is a statistical task. Proposition~\ref{prop:joint_post_distr} follows from the \emph{Causal-ICM} prior and Gaussian likelihood, without further causal assumptions. We now relate the two response surfaces to the causal quantities of interest.

\subsection{Sensitivity Analysis from Observational Data Alone}
\label{sec:obs_only}

When only observational data are available, \emph{Causal-ICM} is most directly comparable to the \gls{osm} of \citet{bonvini2022sensitivityanalysismarginalstructural}, which bounds departures from the observational regression \( \mu(x,a)=\mathbb E[Y\mid X=x,A=a] \). Suppose an unmeasured confounder \( U \) suffices for conditional exchangeability, \( Y(a)\perp A\mid X,U \), and assume SUTVA (\ref{A1:consistency}) and treatment positivity, \( 0<P(A=a\mid X,U)<1 \), for each arm almost surely. Defining \( \mu(x,u,a)=\mathbb E[Y\mid X=x,U=u,A=a] \), the model imposes
\[
|\Delta(a)|\leq\delta,
\qquad
\Delta(a)=\int\{\mu(x,u,a)-\mu(x,a)\}\,dP(x,u).
\]
These assumptions yield the sharp bounds \( \psi(a)-\delta\leq\mathbb E[Y(a)]\leq\psi(a)+\delta \), where \( \psi(a)=\int\mu(x,a)\,dP(x) \). Thus, \( \delta \) controls the maximum absolute average discrepancy on the outcome scale.

In \emph{Causal-ICM}, $\rho$ plays an analogous sensitivity role to $\delta$. With observational data alone, corresponding to \(\mathcal D_{\mathrm{e}}=\emptyset\) in Proposition~\ref{prop:joint_post_distr}, we have information only about the observational surface \(\mu_{\mathrm{o}}\), while the causal surface \(\mu_{\mathrm{e}}\) remains unidentified. Our correlated Gaussian process prior nevertheless links the two surfaces, yielding a posterior over \(\mu_{\mathrm{e}}\) through its assumed dependence on \(\mu_{\mathrm{o}}\). Suppose we observe only \(\mathcal D_{\mathrm{o}}=(y_{\mathrm{o}},X_{\mathrm{o}})\), with \(y_{\mathrm{o}}=\mu_{\mathrm{o}}(X_{\mathrm{o}})+\varepsilon_{\mathrm{o}}\), \(\varepsilon_{\mathrm{o}}\sim\mathcal N(0,\sigma^2 I)\). Under the \emph{Causal-ICM} prior, the likelihood depends on \(\mu_{\mathrm{o}}\) alone,
\begin{equation}
p\!\left(y_{\mathrm{o}}\mid \mu_{\mathrm{e}},\mu_{\mathrm{o}}\right)=p\!\left(y_{\mathrm{o}}\mid \mu_{\mathrm{o}}\right)
=\mathcal N\!\left(y_{\mathrm{o}};\,\mu_{\mathrm{o}}(X_{\mathrm{o}}),\,\sigma^2 I\right),
\end{equation}
so \(y_{\mathrm{o}}\perp \mu_{\mathrm{e}}\mid \mu_{\mathrm{o}}\).

\begin{corollary}[Observational-data-only posterior, \emph{Causal-ICM}]
\label{cor:post_icm_obs_only}
Suppose only $\mathcal D_{\mathrm{o}}$ is observed, and let $\Sigma_{\mathrm{o}} = k(X_{\mathrm{o}},X_{\mathrm{o}}) + \sigma^2 I$. The posterior of the observational surface is $\mu_{\mathrm{o}}(x_*) \mid \mathcal D_{\mathrm{o}} \sim \mathcal N\!\left(m_{\mathrm{o}}(x_*), V_{\mathrm{o}}(x_*)\right)$, where
\[
m_{\mathrm{o}}(x_*) = k(x_*,X_{\mathrm{o}})\,\Sigma_{\mathrm{o}}^{-1} y_{\mathrm{o}}, \qquad
V_{\mathrm{o}}(x_*) = k(x_*,x_*) - k(x_*,X_{\mathrm{o}})\,\Sigma_{\mathrm{o}}^{-1} k(X_{\mathrm{o}},x_*).
\]
Under Causal-ICM, the posterior of the experimental (causal) surface is then
\[
\mu_{\mathrm{e}}(x_*) \mid \mathcal D_{\mathrm{o}} \sim
\mathcal N\!\left(\rho\, m_{\mathrm{o}}(x_*),\; (1-\rho^2)\,k(x_*,x_*) + \rho^2\, V_{\mathrm{o}}(x_*)\right).
\]
\end{corollary}

\begin{proof}
    See Web Appendix~\ref{app:proof:post_icm_obs_only}.
\end{proof}
Corollary~\ref{cor:post_icm_obs_only} has three consequences with parallels to the OSM. First, the causal posterior mean is $\rho\,m_{\mathrm{o}}(x_*)$, where $m_{\mathrm{o}}(x_*)$ is the estimate one would report by treating the observational regression as causal. As \(\rho\to1\) the two coincide (full borrowing), and as \(\rho\to0\) the estimate is pulled to the prior mean. Second, if \(x_*\) lies in the support of the observational covariates, letting \(n_{\mathrm{o}}\to\infty\) sends \(V_{\mathrm{o}}(x_*)\) to some \(c(x_*)\geq0\), with \(c(x_*)=0\) for the RBF kernel used in Section~\ref{sec:simulation-study}, so the causal posterior variance converges to \((1-\rho^2)k(x_*,x_*)+\rho^2c(x_*)\), which stays positive for \(\rho<1\). The term \((1-\rho^2)k(x_*,x_*)\) is the prior variance of the confounding bias given $\mu_{\mathrm{o}}$, which observational data cannot reduce. This is the Bayesian counterpart of the OSM interval \(\psi(a)\pm\delta\), which also does not shrink with sample size, with smaller \(\rho\) behaving like larger \(\delta\). Third, because \(y_{\mathrm{o}} \perp \mu_{\mathrm{e}}\mid \mu_{\mathrm{o}}\), the experimental posterior is obtained entirely through the prior link, i.e.\ \(\pi(\mu_{\mathrm{e}}\mid \mu_{\mathrm{o}},\mathcal D_{\mathrm{o}})=\pi(\mu_{\mathrm{e}}\mid \mu_{\mathrm{o}})\), so that in the absence of trial data, we cannot learn anything about \(\mu_{\mathrm{e}}\) beyond what \(\rho\) reflects from \(\mu_{\mathrm{o}}\). We see below that the availability of trial data reduces this dependence on the prior. 


\subsection{Data Fusion with \gls{rct} Data}
\label{sec:rct-data}



When \gls{rct} data are available, under the trial-specific assumptions (Assumptions~\ref{A1:consistency}--\ref{A3:RCT_A_positivity}) we can identify the experimental causal response surface which, under a study exchangeability assumption (\ref{A5:S_exchangeability}), corresponds with the observational-population causal response surface. 
This means that the discrepancy between the observational and causal regressions becomes identifiable. 
The parameter \(\rho\) governs how strongly the observational regression is allowed to inform the causal estimate. 

The joint posterior in Proposition~\ref{prop:joint_post_distr} immediately yields the marginal posteriors for each study-specific response surface. Writing these marginals separately is useful because it isolates the posterior for \(\mu_{\mathrm{e}}\), the trial-specific causal surface, from the posterior for the observational regression \(\mu_{\mathrm{o}}\).

\begin{corollary}[Posterior distribution of the experimental surface]
\label{cor:marginal_post}
Under \emph{Causal-ICM}, the posterior distribution of the experimental (causal) surface at \(x_*\) is
\[
\mu_{\mathrm{e}}(x_*)\mid\mathcal D\sim\mathcal N\big(m_{\mathrm{e}|\mathcal D}(x_*),\,V_{\mathrm{e}|\mathcal D}(x_*)\big),
\]
where
\begin{align*}
m_{\mathrm{e}|\mathcal D}(x_*)&=m_{\mathrm{e}|\mathcal D_{\mathrm{e}}}(x_*)+\rho\,q(x_*)\Sigma_{\mathrm{eo}}^{-1}r_{\mathrm{o}}, &
V_{\mathrm{e}|\mathcal D}(x_*)&=V_{\mathrm{e}|\mathcal D_{\mathrm{e}}}(x_*)-\rho^2\,q(x_*)\Sigma_{\mathrm{eo}}^{-1}q(x_*)^\top,
\end{align*}
and \(m_{\mathrm{e}|\mathcal D_{\mathrm{e}}}(x_*)=k(x_*,X_{\mathrm{e}})\Sigma_{\mathrm{e}}^{-1}y_{\mathrm{e}}\) and \(V_{\mathrm{e}|\mathcal D_{\mathrm{e}}}(x_*)=k(x_*,x_*)-k'(x_*,x_*)\) are the posterior mean and variance of \(\mu_{\mathrm{e}}(x_*)\) given the trial data \(\mathcal D_{\mathrm{e}}\) alone.
\end{corollary}

\begin{proof}
The marginal of a bivariate normal is normal, with the corresponding entry of the mean vector and the corresponding diagonal entry of the covariance matrix. Applying this to the first component of Proposition~\ref{prop:joint_post_distr} gives the result.
\end{proof}

By Corollary~\ref{cor:marginal_post}, the observational data update the trial-only posterior of the experimental surface through a mean shift and a variance reduction,
\[
m_{\mathrm{e}|\mathcal D}(x_*)=m_{\mathrm{e}|\mathcal D_{\mathrm{e}}}(x_*)+C(x_*;\rho),\qquad V_{\mathrm{e}|\mathcal D}(x_*)=V_{\mathrm{e}|\mathcal D_{\mathrm{e}}}(x_*)-R(x_*;\rho),
\]
\[
C(x_*;\rho)=\rho\,q(x_*)\Sigma_{\mathrm{eo}}^{-1}r_{\mathrm{o}},\qquad R(x_*;\rho)=\rho^2\,q(x_*)\Sigma_{\mathrm{eo}}^{-1}q(x_*)^\top\ge0 .
\]

The shift \(C(x_*;\rho)\) pulls the trial-only posterior mean towards the observational outcomes, in proportion to what the observational data tell us about \(\mu_{\mathrm{e}}(x_*)\) once we condition on trial data. Specifically, \(r_{\mathrm{o}}\) is the part of the observational outcomes that the trial data do not already predict, and it is weighted by its posterior covariance with \(\mu_{\mathrm{e}}(x_*)\) given the trial data. The variance reduction \(R(x_*;\rho)\) is larger when the observational covariates lie close to \(x_*\) and are poorly covered by the trial.

We next show that a large observational sample cannot move the posterior mean of $\mu_{\mathrm{e}}(x_*)$ arbitrarily far from its trial-only value, indicating a level of robustness.

\begin{proposition}[Bounds on the observational shift]
\label{prop:obs_shift_bounds}
Fix \(\mathcal D_{\mathrm{e}}\), \(x_*\), treatment arm \(a\), prior correlation \(\rho\in[0,1]\), and let \(\mathcal H_k\) be the reproducing kernel Hilbert space of \(k\). Let \(P_{\mathrm{o}}^\star\) be the true observational sampling law, and let \(\mu_{\mathrm{o}}(x)=\mathbb E^\star[Y\mid X=x,A=a,S=\mathrm{o}]\) be the observational regression under this law with \(\varepsilon_{\mathrm{o}}=y_{\mathrm{o}}-\mu_{\mathrm{o}}(X_{\mathrm{o}})\).  Suppose the following conditions hold.
\begin{enumerate}[label=(B\arabic*)]
\item The observational regression function $\mu_{\mathrm{o}}$ lies in $\mathcal{H}_k$.
\item The errors \(\varepsilon_{\mathrm{o}}\) are i.i.d.\ with finite variance \(\sigma_\varepsilon^2\), and are independent of \((X_{\mathrm{o}},\mathcal D_{\mathrm{e}})\).
\end{enumerate}
If \(\rho>0\) and \(V_{\mathrm{e}|\mathcal D_{\mathrm{e}}}(x_*)>0\), then
\[
C(x_*;\rho)=O_{P_{\mathrm{o}}^\star}\!\Big(\rho\sqrt{V_{\mathrm{e}|\mathcal D_{\mathrm{e}}}(x_*)}\Big)\qquad\text{as } n_{\mathrm{o}}\to\infty,
\]
conditional on the fixed trial and uniformly over observational designs. If \(\rho=0\) or \(V_{\mathrm{e}|\mathcal D_{\mathrm{e}}}(x_*)=0\), then \(C(x_*;\rho)=0\) for every observational outcome vector.
\end{proposition}

\begin{proof}
The proof is provided in Web Appendix~\ref{app: proof: obs_shift_bounds}.
\end{proof}

Thus, under (B1) and (B2), the shift induced by the observational data is of order $\rho\sqrt{V_{\mathrm{e}\mid\mathcal D_{\mathrm{e}}}(x_*)}$: with the trial data held fixed, letting $n_{\mathrm{o}} \to \infty$ cannot move the posterior mean arbitrarily far from the trial-only estimate. Smaller values of $\rho$ tighten this bound, and at $\rho = 0$ the observational data have no effect.

Observational studies are typically much larger than \glspl{rct}. Under full borrowing, this size imbalance can cause the observational likelihood to dominate the posterior, producing overconfident estimates when the observational regression surface is affected by residual confounding. Having bounded the shift in the posterior mean, we now turn to the posterior variance and show that the reduction in uncertainty attainable from the observational data is likewise limited. The following bound was derived in \citet{dimitriou2025causalicmdatafusionframework}.

\begin{proposition}[Posterior variance bound; \citealp{dimitriou2025causalicmdatafusionframework}]
\label{prop:v_bound}
Let \(V_{\mathrm{e}|\mathcal{D}}(x_*)\) and \(V_{\mathrm{e}|\mathcal{D}_{\mathrm{e}}}(x_*)\) denote the posterior variances of \(\mu_{\mathrm{e}}(x_*)\) given \(\mathcal{D}\) and \(\mathcal{D}_{\mathrm{e}}\) respectively. For \(\rho \in (0,1)\),
\[
V_{\mathrm{e}|\mathcal{D}}(x_*) \geq (1-\rho^2)\, V_{\mathrm{e}|\mathcal{D}_{\mathrm{e}}}(x_*).
\]
\end{proposition}

\begin{proof}
The result follows from the bound on \(R(x_*;\rho)\), established in the proof of Proposition~\ref{prop:obs_shift_bounds}; see Web Appendix~\ref{app: proof: v_bound}.
\end{proof}

Since \(V_{\mathrm{e}| \mathcal D}(x_*) = V_{\mathrm{e}| \mathcal D_{\mathrm{e}}}(x_*) - R(x_*;\rho)\) and \(R(x_*;\rho)\geq 0\), also \(V_{\mathrm{e}| \mathcal D}(x_*) \leq V_{\mathrm{e} | \mathcal D_{\mathrm{e}}}(x_*)\). Together, the two bounds give \((1-\rho^2)V_{\mathrm{e}|\mathcal D_{\mathrm{e}}}(x_*)\leq V_{\mathrm{e}|\mathcal D}(x_*)\leq V_{\mathrm{e}|\mathcal D_{\mathrm{e}}}(x_*)\). Thus, adding observational data can reduce the posterior variance of the experimental surface by at most \(\rho^2\) relative to the trial-only posterior variance.
At \(\rho=1\), the lower bound is \(0\); at \(\rho=0\), it equals \(V_{\mathrm{e}|\mathcal D_{\mathrm{e}}}(x_*)\). For \(\rho \in (0,1)\), the bound is independent of \(n_{\mathrm{o}}\): posterior variance under \(\mathcal D\) does not converge to \(0\) as \(n_{\mathrm{o}}\to\infty\).

\subsection{Identification of the Confounding Bias} \label{sec:bias_identification}

Under \ref{A1:consistency}--\ref{A6:S_positivity} the arm-specific confounding bias is $d(x,a)=\mu_{\mathrm{o}}(x,a)-\mu_{\mathrm{e}}(x,a)$. In a single observational study this discrepancy is unidentified and can only be bounded, as in the \gls{osm} through $\delta$. With trial data and under our assumptions, it is identifiable, and under \emph{Causal-ICM} it has a
closed-form posterior. For each arm $a$, let $\Delta_{\mathrm{OSM}}(a)=\int\{\mu_{\mathrm{o}}(x,u,a)-\mu_{\mathrm{o}}(x,a)\}\,dP(x,u\mid S=\mathrm{o})$.

\begin{proposition}[Identification and posterior of the confounding bias] \label{prop:bias}
    Fix $a\in\{0,1\}$. Suppose Assumptions~\ref{A1:consistency}--\ref{A6:S_positivity} hold. Then
    \[
    \Delta_{\mathrm{OSM}}(a)=-\int d(x,a)\,dP(x\mid S=\mathrm{o}).
    \]
    Under \emph{Causal-ICM}, for any test point $x_*$,
    \[
    d(x_*,a)\mid\mathcal{D}\sim\mathcal{N}\Big(m_{\mathrm{o}|\mathcal{D}}(x_*)-m_{\mathrm{e}|\mathcal{D}}(x_*),\;
    V_{\mathrm{o}|\mathcal{D}}(x_*)+V_{\mathrm{e}|\mathcal{D}}(x_*)-2C_{\mathrm{eo}|\mathcal{D}}(x_*)\Big).
    \]
\end{proposition}
\begin{proof}
The proof is provided in Web Appendix~\ref{app:proof:osm-bias-identification}.
\end{proof}

Trial data make the observational confounding bias identifiable, unlike in the observational-only \gls{osm}, where \(\Delta_{\mathrm{OSM}}(a)\) is bounded by \(\delta\) rather than learned from the data \citep{bonvini2022sensitivityanalysismarginalstructural}.
The parameter \(\rho\) controls the prior dependence between the surfaces: \(\rho=1\) forces \(\mu_{\mathrm{o}}=\mu_{\mathrm{e}}\) and hence \(d\equiv0\), whereas smaller values allow greater discrepancies. Under arm-specific independence, the posterior of \(D(x_*)=d(x_*,1)-d(x_*,0)\) is Gaussian, with mean equal to the difference of the arm-specific posterior means and variance equal to the sum of their posterior variances.

\section{Numerical Results}
 
\subsection{Simulation Study}
\label{sec:simulation-study}
 
To illustrate how \emph{Causal-ICM} can be used in a sensitivity analysis, we simulate an observational study and a randomised trial from a common data-generating process (DGP) with a known, unmeasured confounder in the observational study.
 
Let $X, U \sim \mathcal{N}(0,1)$ be independent covariates, with $X$ observed in both studies and $U$ observed in neither. Treatment assignment and outcomes are generated as
\begin{align*}
A \mid X, U, S=\mathrm{o} &\sim \mathrm{Bernoulli}\big(\operatorname{expit}(\alpha_0 + \alpha_x X + \alpha_u U)\big), \\
A \mid X, U, S=\mathrm{e} &\sim \mathrm{Bernoulli}(0.5), \\
Y &= \beta_0 + \tau A + \beta_x X + \beta_u U + \varepsilon, \qquad \varepsilon \sim \mathcal{N}(0,\sigma_Y^2),
\end{align*}
where $S \in \{\mathrm{o},\mathrm{e}\}$ indexes the observational and experimental studies respectively. In the observational study, treatment is confounded through $U$, while in the RCT treatment is assigned independently of $(X,U)$ by design, satisfying Assumption~\ref{A2:RCT_A_exchangeability}. We use $(\alpha_0,\alpha_x,\alpha_u) = (1.0,\,0.8,\,1.0)$ and $(\beta_0,\tau,\beta_x,\beta_u,\sigma_Y) = (0,\,0.2,\,1.0,\,1.0,\,1.0)$, with $n_{\mathrm{o}} = 5000$ observational units and $n_{\mathrm{e}} = 50$ RCT units. The outcome is centred by subtracting its pooled sample mean across both studies before analysis, so that all reported levels are deviations from the overall mean; the ATE, being a contrast, is unaffected by this transformation.

We select the hyperparameters of the RBF kernel for the \emph{Causal-ICM} models in the same way, using the observational data for the observational-only posterior and the RCT data for the data-fusion posterior. Covariates are standardised with the RCT mean and standard deviation; the kernel length scale is set using the median pairwise distance between covariate values, the signal variance to the empirical outcome variance, and the noise variance is estimated from the residuals of a linear model. For \textit{Causal-ICM} under data fusion, the \gls{gp} prior means are the RCT arm means. For the observational-only \textit{Causal-ICM}, the prior mean for both arms is the mean outcome over all observational units (treated and control combined); as $\rho\to0$ each arm's posterior mean reverts to the same prior mean, so the posterior mean for the \gls{ate} centres around zero.
 
We compare three sensitivity analyses: the OSM of \citet{bonvini2022sensitivityanalysismarginalstructural}, the \emph{Causal-ICM} posterior using observational data alone%
, and the \emph{Causal-ICM} posterior using the RCT and observational data jointly, i.e.\ data fusion
, for $\mathbb{E}[Y(0)]$, $\mathbb{E}[Y(1)]$, and the ATE $\tau = \mathbb{E}[Y(1)] - \mathbb{E}[Y(0)]$. 

\begin{figure}[htbp]
    \setstretch{1}
    \centering
    \includegraphics[width=\textwidth]{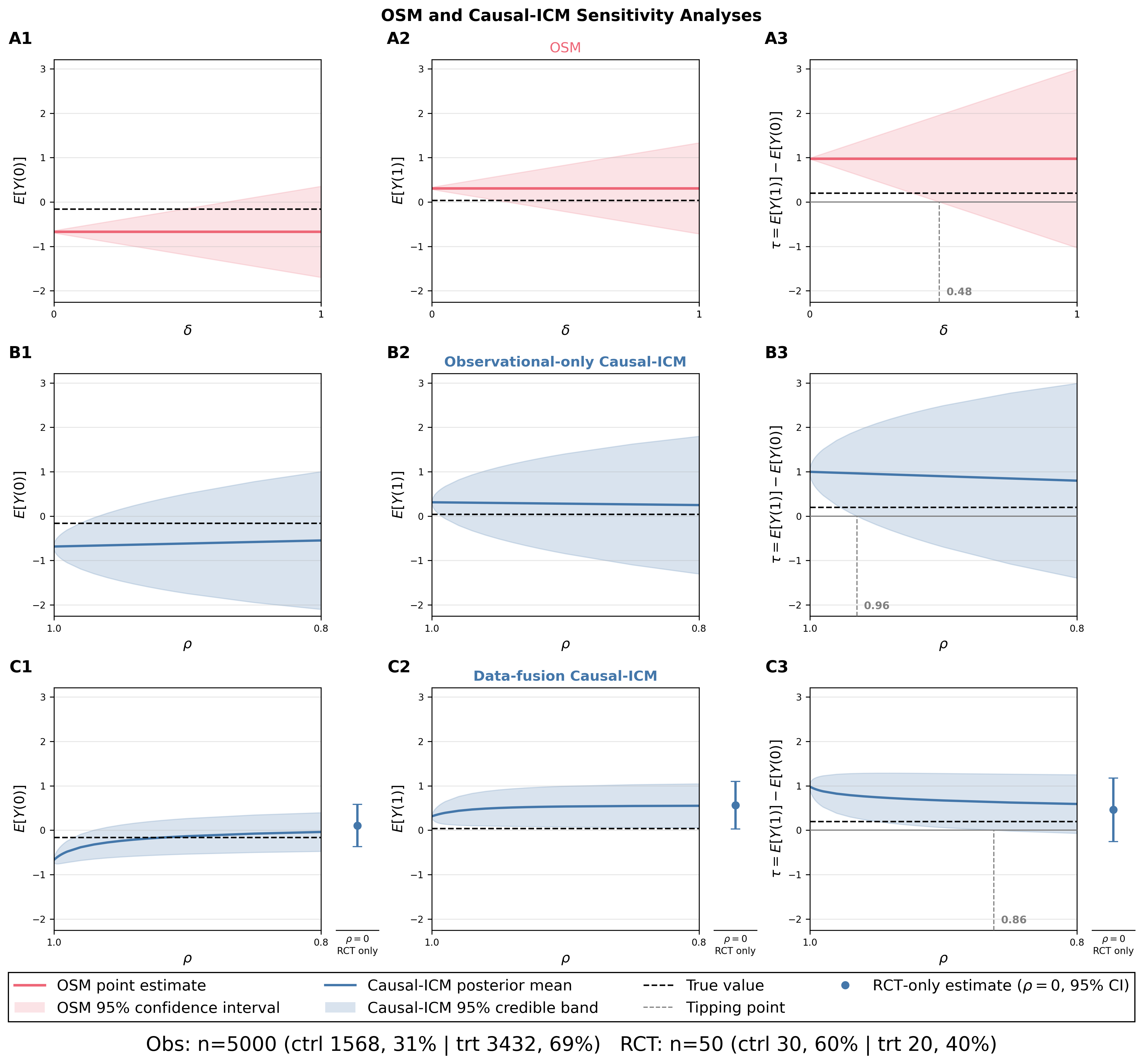}
    \caption{OSM and \emph{Causal-ICM} sensitivity analyses for $\mathbb{E}[Y(0)]$ (column 1), $\mathbb{E}[Y(1)]$ (column 2), and the ATE $\tau=\mathbb{E}[Y(1)]-\mathbb{E}[Y(0)]$ (column 3). Row A shows the OSM analysis of \citet{bonvini2022sensitivityanalysismarginalstructural}, as a function of the sensitivity parameter $\delta$. Row B shows the \emph{Causal-ICM} posterior using observational data alone, as a function of the prior correlation $\rho$. Row C shows the \emph{Causal-ICM} posterior using the RCT and observational data jointly (data fusion), with the RCT-only estimate ($\rho=0$, point with 95\% CI) shown alongside each panel for reference. The dashed black line marks the true value; the dashed grey vertical line marks the tipping point at which the lower bound of the 95\% interval for the ATE crosses zero ($\delta^\ast\approx0.48$, panel A3; $\rho^\ast\approx0.96$, panel B3; $\rho^\ast\approx0.86$, panel C3). Relative observational and RCT treatment/control sample sizes are reported below the figure.}
    \label{fig:sim-all-methods}
\end{figure}

Figure~\ref{fig:sim-all-methods} compares the three approaches. The \gls{osm} point estimate (row A) does not change with $\delta$, because the \gls{osm} widens bounds around a fixed point estimate. The resulting `tipping point', where the \gls{ate}'s lower bound crosses zero, is $\delta^\ast \approx 0.48$ (panel A3). Under \emph{Causal-ICM}, by contrast, both the point estimate and the interval change with $\rho$, because $\rho$ changes the posterior itself. With observational data alone (row B), the \gls{ate}'s tipping point is $\rho^\ast \approx 0.96$ (panel B3), close to full pooling ($\rho=1$). Introducing \gls{rct} data results in the joint posterior mean decreasing further as $\rho$ decreases, from $\approx1.0$ at $\rho=1$ to $\approx0.6$ at $\rho=0.80$, and the tipping point moves to $\rho^\ast \approx 0.86$ (panel C3). This is more robust than the observational-only case, consistent with the robustness properties established in Propositions~\ref{prop:obs_shift_bounds} and~\ref{prop:v_bound}, and requires a larger relaxation of the treatment exchangeability assumption in the observational study before the conclusion changes. Panels C1--C3 also show the RCT-only estimate ($\rho=0$, point with 95\% CI) for reference; the data-fusion band contains the true \gls{ate} only for $\rho\leq0.95$.

 
\subsection{Application to PIONEER 6 and Real-World Data}
\label{sec:pioneer6-application}
 
We now apply a \textit{Causal-ICM} sensitivity analysis to estimation of the effect of oral semaglutide on HbA1c in patients with type 2 diabetes using data obtained by combining synthetic versions of the PIONEER~6 randomised trial \citep{husain2019oral} and a large US health claims database. 
PIONEER~6 was a randomised, double-blind cardiovascular safety trial that also recorded HbA1c after one year of treatment \citep{husain2019oral}. The claims database ($S=\mathrm{o}$) provides a larger observational sample of patients in routine care and may be affected by residual confounding and by differences from the trial population. 

The synthetic datasets were generated to reproduce key characteristics of the real datasets, using the frengression method of \citet{yang2026frugalflexiblefaithfulcausal}. Frengression is a deep generative framework that models the relationships between variables around a specific causal effect to accurately estimate, faithfully simulate, and directly sample from complex, time-varying interventional data. The outcome for the trial ($S=\mathrm{e}$) is HbA1c at one-year follow-up. To control for measured confounding, we adjust for baseline covariates available in both sources: sex, age, obesity history, diabetes, antihypertensive and lipid-lowering medication use, baseline HbA1c, HDL and LDL cholesterol, and serum creatinine. The outcome, covariates and prior means are processed as in Section~\ref{sec:simulation-study}, and the target population is the trial population. 

Full pooling ($\rho=1$) treats the claims-based HbA1c regressions as equal to the trial-identified response surfaces. We assess how far $\rho$ can be lowered from 1 before the conclusion changes, that is, where the tipping point lies (Web Appendix~\ref{app:tipping-point}).
For the observational-only \emph{Causal-ICM} posterior, we set the \gls{gp} hyperparameters by fitting two independent \glspl{gp}, one per treatment arm, to the observational data alone, optimising the marginal likelihood. Given the size of the claims database, we use inducing points \citep{titsias2009variational} to speed up this optimisation. 
For the data fusion \emph{Causal-ICM} posterior, we fit the same two \glspl{gp} to the RCT data instead, using the full trial data since the trial is comparatively small. 
This ties the hyperparameters to the randomised data rather than to the larger and possibly confounded claims data.
For the subgroup analyses, we repeat this step within each subgroup, so that length-scales and noise variance can differ between subgroups.
 
We report the \emph{Causal-ICM} data-fusion posterior for the full sample and for two subgroups defined by baseline HbA1c, using the clinical threshold of $5.7\%$ that is the threshold between non-diabetic and prediabetic patients. These subgroup analyses were not prespecified and are interpreted as exploratory. Among participants with established type 2 diabetes, this distinguishes baseline glycaemic levels rather than diabetes status.

\begin{figure}[htbp]
    \setstretch{1}
    \centering
    \includegraphics[width=0.94\textwidth]{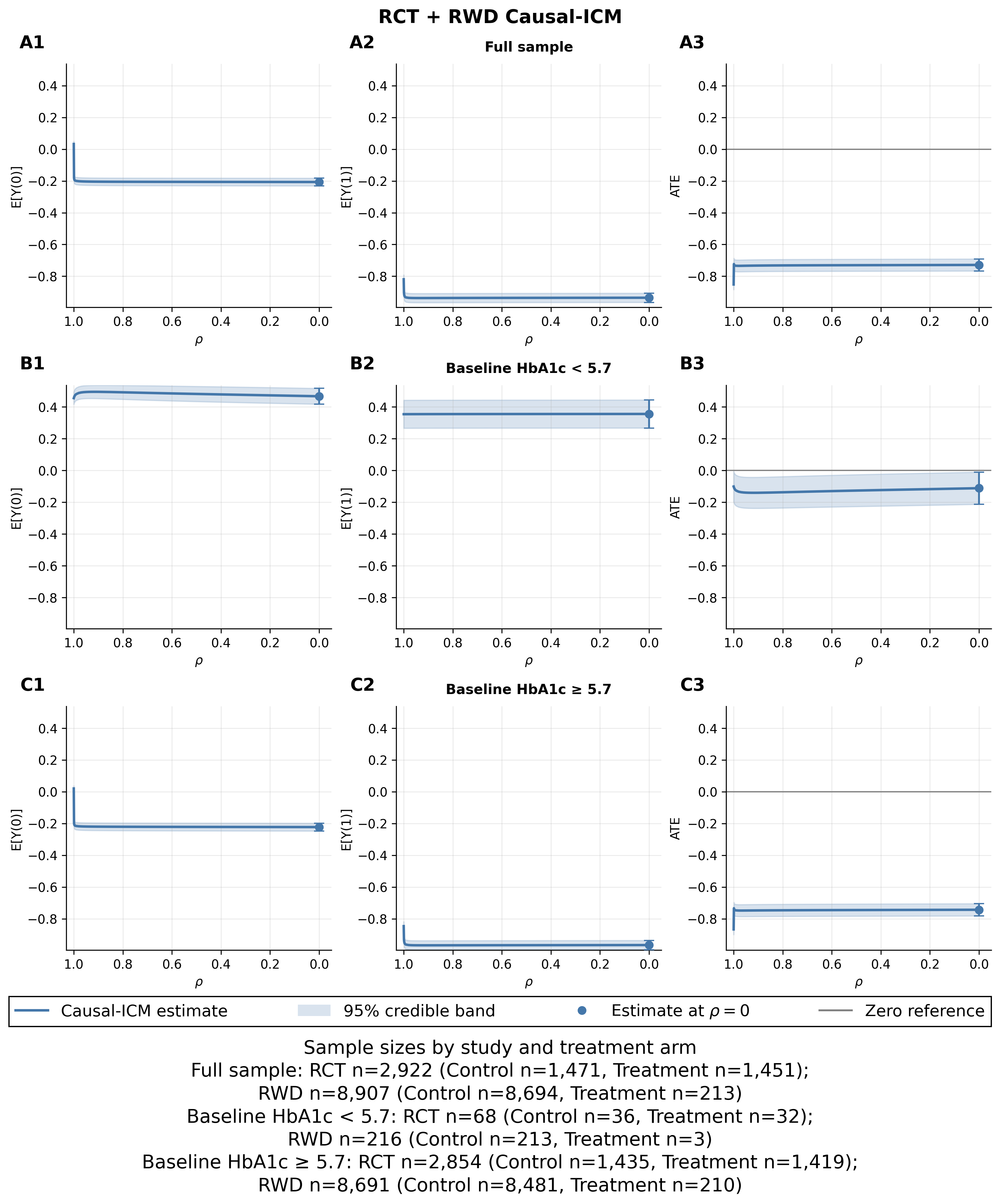}
    \caption{\emph{Causal-ICM} data-fusion posterior using the PIONEER~6 trial and claims data, for $\mathbb{E}[Y(0)]$, $\mathbb{E}[Y(1)]$ and the ATE (columns 1--3), plotted against $\rho$, for the full sample (row A) and baseline HbA1c $<5.7$ (row B) and $\geq5.7$ (row C). The point with error bar at $\rho=0$ gives the RCT-only estimate with its 95\% CI; the grey line in the ATE panels marks zero.}
    \label{fig:pioneer6}
\end{figure}

Figure~\ref{fig:pioneer6} shows that the 95\% credible band of the ATE excludes zero for every $\rho\in[0,1]$, in the full sample and in both subgroups, so there is no tipping point. In the full sample (row A), the joint posterior mean of the ATE moves from approximately $-0.85$ at full pooling ($\rho=1$) to $-0.73$ at the RCT-only endpoint ($\rho=0$), the latter consistent with the original PIONEER~6 trial \citep{husain2019oral}; moving towards full pooling shifts the point estimate further from zero and narrows the credible band, so the claims data strengthen the trial's conclusion. Patients with baseline HbA1c of at least $5.7$ (row C) drive this result, with the ATE moving from approximately $-0.87$ to $-0.74$. Among patients with baseline HbA1c below $5.7$ (row B), the effect is much smaller in magnitude ($-0.10$ at $\rho=1$ and $-0.11$ at $\rho=0$) and the band comes close to including zero, reflecting both a smaller true effect and fewer patients in this subgroup.

In all three rows, the posterior mean and credible band change sharply between $\rho=1$ and $\rho=0.99$ and are nearly flat below that, so the conclusions are insensitive to departures from the no unmeasured confounding assumption in the observational study. Real-world data alone tell a mixed story. While the OSM intervals include zero in both subgroups, observational-only \emph{Causal-ICM} is uninformative for patients with baseline HbA1c $\geq5.7$ yet estimates a negative effect for those with HbA1c $<5.7$ when $\rho>0.95$. Adding the RCT identifies the effect for the $\geq5.7$ subgroup and recovers the full-sample result. For further details see Web Appendix~\ref{app:pioneer6-rwd-only}. 

\section{Discussion} \label{sec:discussion}

We proposed a Bayesian nonparametric sensitivity analysis that does not assume the observational outcome regression and the causal response surface are the same function; we instead model them as two related functions with a joint prior, and the prior correlation $\rho$ quantifies how close they are. The joint prior induces a prior on the confounding bias whose scale is set by $\rho$, so each value of $\rho$ gives a posterior for the causal response surface rather than a bound. With observational data alone, this posterior reflects uncertainty about the bias as well as sampling variability, and its variance does not vanish as $n_{\mathrm{o}}\to\infty$ (Corollary~\ref{cor:post_icm_obs_only}). With trial data added, full pooling ($\rho=1$) and the trial-only analysis ($\rho=0$) become the two endpoints of the same analysis. The confounding bias has a closed-form posterior (Proposition~\ref{prop:bias}), and the influence of the observational data on the trial-only posterior is limited by $\rho$ (Proposition~\ref{prop:v_bound}).

The PIONEER~6 application shows how the method can be used in practice. We emphasise that the analysis uses synthetic versions of the trial and the claims database, so it illustrates the method rather than providing clinical evidence on oral semaglutide. In the full sample, the trial's conclusion holds at every value of $\rho$, and the claims data serve to add precision, mainly when $\rho$ is close to 1. More generally, we recommend reporting the whole curve in $\rho$, from the trial-only analysis ($\rho=0$) to full pooling ($\rho=1$), together with any tipping point, rather than deciding whether to pool. Read this way, the curve in $\rho$ shows how the conclusion depends on the degree of borrowing from the observational data.

Several limitations suggest extensions. First, $\rho$ is a single scalar shared across covariate values and treatment arms; allowing $\rho(x)$ or arm-specific correlations would allow the degree of borrowing to vary across the covariate space. Second, interpreting the difference between the observational and trial outcome regressions as confounding bias relies on Assumption~\ref{A5:S_exchangeability}. If \ref{A5:S_exchangeability} fails, this difference also absorbs differences between the studies' causal response surfaces, which the current model does not distinguish from confounding. Incorporating a transportability component as in \citet{lanners2025datafusionpartialidentification} could separate the two. Third, although the induced prior on the bias gives $\rho$ an outcome-scale interpretation, that interpretation depends on the kernel variance; calibrating $\rho$ against the identified bias $D(x)$ offers an alternative. Fourth, the kernel and noise hyperparameters are fixed at plug-in values, so the credible bands are conditional on these as well as $\rho$. A fully Bayesian approach would propagate their uncertainty. Fifth, our closed-form results rely on continuous outcomes with Gaussian noise. Binary and survival outcomes would require alternative likelihoods and corresponding extensions to posterior inference. Finally, our approach places sensitivity on the outcome regressions, but the same idea could be applied to treatment assignment. A Bayesian counterpart of the Marginal Sensitivity Model~\citep{tan2005MSM} would replace the bound $\Lambda$ on the odds ratio between propensity scores with and without $U$ by a prior on that ratio. With trial data, $\Lambda$ may be partially identified, in which case it could be constrained by data rather than fixed in advance.

\section*{Acknowledgments}
E.D. was funded by the Engineering and Physical Sciences Research Council [grant number EP/W524335/1] and by Novo Nordisk A/S, which also provided the data. 
E.F. receives funding from the Research Grants Council of Hong Kong through the
Early Career Scheme (Grant 27304424) and the General Research Fund (Grant 17306925).
J.M.T. is employed by Novo Nordisk A/S. 
K.D-O was funded by a Royal Society-Wellcome Trust Sir Henry Dale fellowship (218554/Z/19/Z). 
B.L. received research funding from Novo Nordisk A/S. Claude (Anthropic) was used for language editing and code.

\section*{Data Availability}
The data underlying the synthetic data analysis cannot be shared for commercial reasons.

\printbibliography

@article{berger1994overview,
  author  = {Berger, James O.},
  title   = {An overview of robust {B}ayesian analysis},
  journal = {Test},
  year    = {1994},
  volume  = {3},
  number  = {1},
  pages   = {5--124},
  doi     = {10.1007/BF02562676}
}

@article{gustafson2010bayesian,
  author  = {Gustafson, Paul},
  title   = {Bayesian Inference for Partially Identified Models},
  journal = {The International Journal of Biostatistics},
  year    = {2010},
  volume  = {6},
  number  = {2},
  pages   = {Article 17},
  doi     = {10.2202/1557-4679.1206}
}

@article{tan2005MSM,
  ISSN = {01621459},
  author = {Zhiqiang Tan},
  journal = {Journal of the American Statistical Association},
  number = {476},
  pages = {1619--1637},
  publisher = {[American Statistical Association, Taylor \& Francis, Ltd.]},
  title = {A Distributional Approach for Causal Inference Using Propensity Scores},
  volume = {101},
  year = {2006},
  doi = {10.1198/016214506000000023},
  url = {https://doi.org/10.1198/016214506000000023}
}

@article{vanderweeleEvalue,
  title = {Sensitivity Analysis in Observational Research: Introducing the {E}-Value},
  author = {Tyler J VanderWeele and Peng Ding},
  journal = {Annals of Internal Medicine},
  volume = {167},
  number = {4},
  pages = {268--274},
  year = {2017},
  doi = {10.7326/M16-2607},
  note = {PMID: 28693043},
  url = {https://doi.org/10.7326/M16-2607}
}

@article{cinelli2019,
  author = {Cinelli, Carlos and Hazlett, Chad},
  title = {Making Sense of Sensitivity: Extending Omitted Variable Bias},
  journal = {Journal of the Royal Statistical Society Series B: Statistical Methodology},
  volume = {82},
  number = {1},
  pages = {39--67},
  year = {2020},
  month = {2},
  issn = {1369-7412},
  doi = {10.1111/rssb.12348},
  url = {https://doi.org/10.1111/rssb.12348}
}

@article{franks2019flexiblesensitivityanalysisobservational,
  title = {Flexible Sensitivity Analysis for Observational Studies Without Observable Implications},
  author = {Alexander M. Franks and Alexander D'Amour and Avi Feller},
  journal = {Journal of the American Statistical Association},
  volume = {115},
  number = {532},
  pages = {1730--1746},
  year = {2020},
  doi = {10.1080/01621459.2019.1604369},
  url = {https://doi.org/10.1080/01621459.2019.1604369},
  eprint = {1809.00399},
  archivePrefix = {arXiv},
  primaryClass = {stat.ME}
}

@article{nabi2024,
  author = {Nabi, Razieh and Bonvini, Matteo and Kennedy, Edward H and Huang, Ming-Yueh and Smid, Marcela and Scharfstein, Daniel O},
  title = {Semiparametric sensitivity analysis: unmeasured confounding in observational studies},
  journal = {Biometrics},
  volume = {80},
  number = {4},
  pages = {ujae106},
  year = {2024},
  month = {10},
  issn = {0006-341X},
  doi = {10.1093/biomtc/ujae106},
  url = {https://doi.org/10.1093/biomtc/ujae106}
}

@article{zhao2018sensitivityanalysisinverseprobability,
  title = {Sensitivity Analysis for Inverse Probability Weighting Estimators via the Percentile Bootstrap},
  author = {Qingyuan Zhao and Dylan S. Small and Bhaswar B. Bhattacharya},
  journal = {Journal of the Royal Statistical Society Series B: Statistical Methodology},
  volume = {81},
  number = {4},
  pages = {735--761},
  year = {2019},
  doi = {10.1111/rssb.12327},
  url = {https://doi.org/10.1111/rssb.12327},
  eprint = {1711.11286},
  archivePrefix = {arXiv},
  primaryClass = {stat.ME}
}

@inproceedings{lanners2025datafusionpartialidentification,
  title = {Data Fusion for Partial Identification of Causal Effects},
  author = {Quinn Lanners and Cynthia Rudin and Alexander Volfovsky and Harsh Parikh},
  year = {2025},
  doi = {10.52202/085713-5038},
  url = {https://doi.org/10.52202/085713-5038},
  booktitle = {Advances in Neural Information Processing Systems},
  volume = {38},
  pages = {150638--150702},
  publisher = {Curran Associates, Inc.}
}

@inbook{Rosenbaum2002,
  author = {Rosenbaum, Paul R.},
  title = {Sensitivity to Hidden Bias},
  bookTitle = {Observational Studies},
  year = {2002},
  publisher = {Springer New York},
  address = {New York, NY},
  pages = {105--170},
  isbn = {978-1-4757-3692-2},
  doi = {10.1007/978-1-4757-3692-2_4},
  url = {https://doi.org/10.1007/978-1-4757-3692-2_4}
}

@article{cornfield1959smoking,
  title = {Smoking and Lung Cancer: Recent Evidence and a Discussion of Some Questions},
  author = {Cornfield, Jerome and Haenszel, William and Hammond, E. Cuyler and Lilienfeld, Abraham M. and Shimkin, Michael B. and Wynder, Ernst L.},
  journal = {Journal of the National Cancer Institute},
  volume = {22},
  number = {1},
  pages = {173--203},
  year = {1959},
  publisher = {Oxford University Press},
  doi = {10.1093/jnci/22.1.173},
  url = {https://doi.org/10.1093/jnci/22.1.173}
}

@inproceedings{alaa2017bayesianinferenceindividualizedtreatment,
  title = {Bayesian Inference of Individualized Treatment Effects using Multi-task {G}aussian Processes},
  author = {Ahmed M. Alaa and Mihaela van der Schaar},
  year = {2017},
  url = {https://proceedings.neurips.cc/paper/2017/hash/6a508a60aa3bf9510ea6acb021c94b48-Abstract.html},
  booktitle = {Advances in Neural Information Processing Systems},
  volume = {30},
  pages = {3424--3432},
  publisher = {Curran Associates, Inc.}
}

@inproceedings{witty2020causalinferenceusinggaussian,
  title = {Causal Inference using {G}aussian Processes with Structured Latent Confounders},
  author = {Sam Witty and Kenta Takatsu and David Jensen and Vikash Mansinghka},
  year = {2020},
  url = {https://proceedings.mlr.press/v119/witty20a.html},
  booktitle = {Proceedings of the 37th International Conference on Machine Learning},
  series = {Proceedings of Machine Learning Research},
  volume = {119},
  pages = {10313--10323},
  publisher = {PMLR}
}

@inproceedings{kallus2018removinghiddenconfoundingexperimental,
  title = {Removing Hidden Confounding by Experimental Grounding},
  author = {Nathan Kallus and Aahlad Manas Puli and Uri Shalit},
  year = {2018},
  url = {https://dl.acm.org/doi/10.5555/3327546.3327747},
  booktitle = {Advances in Neural Information Processing Systems},
  volume = {31},
  pages = {10888--10897},
  publisher = {Curran Associates, Inc.}
}

@inproceedings{yang2022integrative,
  title = {Integrative $R$-learner of heterogeneous treatment effects combining experimental and observational studies},
  author = {Wu, Lili and Yang, Shu},
  booktitle = {Proceedings of the First Conference on Causal Learning and Reasoning},
  pages = {904--926},
  year = {2022},
  editor = {Schölkopf, Bernhard and Uhler, Caroline and Zhang, Kun},
  volume = {177},
  series = {Proceedings of Machine Learning Research},
  month = {11--13 Apr},
  publisher = {PMLR},
  doi = {},
  url = {https://proceedings.mlr.press/v177/wu22a.html}
}

@article{yang2022elasticintegrativeanalysisrandomized,
  title = {Elastic Integrative Analysis of Randomised Trial and Real-World Data for Treatment Heterogeneity Estimation},
  author = {Shu Yang and Chenyin Gao and Donglin Zeng and Xiaofei Wang},
  journal = {Journal of the Royal Statistical Society Series B: Statistical Methodology},
  volume = {85},
  number = {3},
  pages = {575--596},
  year = {2023},
  doi = {10.1093/jrsssb/qkad017},
  url = {https://doi.org/10.1093/jrsssb/qkad017},
  eprint = {2005.10579},
  archivePrefix = {arXiv},
  primaryClass = {stat.ME}
}

@article{lin2024combiningexperimentalobservationaldata,
  title = {Combining Experimental and Observational Data Through a Power Likelihood},
  author = {Xi Lin and Jens Magelund Tarp and Robin J. Evans},
  journal = {Biometrics},
  volume = {81},
  number = {1},
  pages = {ujaf008},
  year = {2025},
  doi = {10.1093/biomtc/ujaf008},
  url = {https://doi.org/10.1093/biomtc/ujaf008},
  eprint = {2304.02339},
  archivePrefix = {arXiv},
  primaryClass = {stat.ME}
}

@article{colnet2023causalinferencemethodscombining,
  title = {Causal Inference Methods for Combining Randomized Trials and Observational Studies: A Review},
  author = {Bénédicte Colnet and Imke Mayer and Guanhua Chen and Awa Dieng and Ruohong Li and Gaël Varoquaux and Jean-Philippe Vert and Julie Josse and Shu Yang},
  journal = {Statistical Science},
  volume = {39},
  number = {1},
  pages = {165--191},
  year = {2024},
  doi = {10.1214/23-STS889},
  url = {https://doi.org/10.1214/23-STS889},
  eprint = {2011.08047},
  archivePrefix = {arXiv},
  primaryClass = {stat.ME}
}

@misc{lin2024datafusionefficiencygain,
  title = {Data Fusion for Efficiency Gain in {ATE} Estimation: A Practical Review with Simulations},
  author = {Xi Lin and Jens Magelund Tarp and Robin J. Evans},
  year = {2024},
  eprint = {2407.01186},
  archivePrefix = {arXiv},
  primaryClass = {stat.ME},
  doi = {},
  url = {https://arxiv.org/abs/2407.01186}
}

@inproceedings{dimitriou2025causalicmdatafusionframework,
  title = {Causal-ICM: A Data Fusion Framework For Heterogeneous Treatment Effect Estimation With Multi-Task {G}aussian Processes},
  author = {Evangelos Dimitriou and Edwin Fong and Jens Magelund Tarp and Karla Diaz-Ordaz and Brieuc Lehmann},
  year = {2026},
  url = {https://proceedings.mlr.press/v323/dimitriou26a.html},
  booktitle = {Proceedings of the Fifth Conference on Causal Learning and Reasoning},
  series = {Proceedings of Machine Learning Research},
  volume = {323},
  pages = {248--276},
  publisher = {PMLR}
}

@misc{bonvini2022sensitivityanalysismarginalstructural,
  title = {Sensitivity Analysis for Marginal Structural Models},
  author = {Matteo Bonvini and Edward Kennedy and Valerie Ventura and Larry Wasserman},
  year = {2022},
  eprint = {2210.04681},
  archivePrefix = {arXiv},
  primaryClass = {stat.ME},
  doi = {},
  url = {https://arxiv.org/abs/2210.04681}
}

@book{hernan2020causal,
  title = {Causal Inference: What If},
  author = {Hernán, M. A. and Robins, J. M.},
  isbn = {9781420076165},
  series = {Chapman \& Hall/CRC Monographs on Statistics \& Applied Probability},
  year = {2020},
  publisher = {Chapman \& Hall/CRC},
  url = {https://miguelhernan.org/whatifbook}
}

@article{Stuart2011,
  author = {Stuart, Elizabeth A. and Cole, Stephen R. and Bradshaw, Catherine P. and Leaf, Philip J.},
  title = {The Use of Propensity Scores to Assess the Generalizability of Results from Randomized Trials},
  journal = {Journal of the Royal Statistical Society: Series A},
  volume = {174},
  number = {2},
  pages = {369--386},
  year = {2011},
  doi = {10.1111/j.1467-985X.2010.00673.x},
  url = {https://doi.org/10.1111/j.1467-985X.2010.00673.x}
}

@misc{yang2026datafusiondistributionalequivalence,
  title = {Data Fusion with Distributional Equivalence Test-then-pool},
  author = {Yang, Linying and Liu, Xing and Evans, Robin J.},
  year = {2026},
  eprint = {2603.11867},
  archivePrefix = {arXiv},
  primaryClass = {stat.ME},
  url = {https://arxiv.org/abs/2603.11867}
}

@article{zhang2025enhancedmarginalsensitivitymodel,
  title = {Enhanced Marginal Sensitivity Model and Bounds},
  author = {Yi Zhang and Wenfu Xu and Zhiqiang Tan},
  year = {2026},
  doi = {10.1080/01621459.2026.2699465},
  url = {https://doi.org/10.1080/01621459.2026.2699465},
  journal = {Journal of the American Statistical Association},
  pages = {1--13},
  note = {Advance online publication}
}

@article{comment2022bayesian,
  title = {Bayesian data fusion: Probabilistic sensitivity analysis for unmeasured confounding using informative priors based on secondary data},
  author = {Comment, Leah and Coull, Brent A. and Zigler, Corwin and Valeri, Linda},
  journal = {Biometrics},
  volume = {78},
  number = {2},
  pages = {730--741},
  year = {2022},
  month = {June},
  doi = {10.1111/biom.13436},
  url = {https://doi.org/10.1111/biom.13436}
}

@article{dhawan2026bayesian,
  title = {Bayesian Sensitivity of Causal Inference Estimators under Evidence-Based Priors},
  author = {Dhawan, Nikita and Shen, Daniel and Cotta, Leonardo and Maddison, Chris J.},
  journal = {Transactions on Machine Learning Research},
  year = {2026},
  url = {https://openreview.net/forum?id=0zqt85NUyK},
  issn = {2835-8856}
}

@inproceedings{fawkes2025hardnessvalidatingobservationalstudies,
  title = {The Hardness of Validating Observational Studies with Experimental Data},
  author = {Jake Fawkes and Michael O'Riordan and Athanasios Vlontzos and Oriol Corcoll and Ciarán Mark Gilligan-Lee},
  year = {2025},
  url = {https://proceedings.mlr.press/v258/fawkes25b.html},
  booktitle = {Proceedings of The 28th International Conference on Artificial Intelligence and Statistics},
  series = {Proceedings of Machine Learning Research},
  volume = {258},
  pages = {1819--1827},
  publisher = {PMLR}
}

@misc{yang2026frugalflexiblefaithfulcausal,
      title={Frugal, Flexible, Faithful: Causal Data Simulation via Frengression}, 
      author={Linying Yang and Jens Magelund Tarp and Robin J. Evans and Xinwei Shen},
      year={2026},
      eprint={2508.01018},
      archivePrefix={arXiv},
      primaryClass={stat.ME},
      url={https://arxiv.org/abs/2508.01018}, 
}

@article{husain2019oral,
  title={Oral Semaglutide and Cardiovascular Outcomes in Patients with Type 2 Diabetes},
  author={Husain, Mansoor and Birkenfeld, Andreas L. and Donsmark, Morten and Dungan, Kathleen and Eliaschewitz, Freddy G. and Franco, Denise R. and Jeppesen, Ole K. and Lingvay, Ildiko and Mosenzon, Ofri and Pedersen, Sue D. and Tack, Cees J. and Thomsen, Mette and Vilsb{\o}ll, Tina and Warren, Mark L. and Bain, Stephen C.},
  journal={New England Journal of Medicine},
  volume={381},
  number={9},
  pages={841--851},
  year={2019},
  doi={10.1056/NEJMoa1901118},
  pmid={31189517}
}

@article{mccandless2007bayesian,
  author = {McCandless, Lawrence C. and Gustafson, Paul and Levy, Adrian R.},
  title = {Bayesian sensitivity analysis for unmeasured confounding in observational studies},
  journal = {Statistics in Medicine},
  volume = {26},
  number = {11},
  pages = {2331--2347},
  year = {2007},
  publisher = {Wiley Online Library},
  doi = {10.1002/sim.2711},
  url = {https://doi.org/10.1002/sim.2711}
}

@article{ung2025generalizing,
  author    = {Ung, Lawson and VanderWeele, Tyler J. and Dahabreh, Issa J.},
  title     = {Generalizing and Transporting Causal Inferences from Randomized Trials in the Presence of Trial Engagement Effects},
  journal   = {Epidemiology},
  volume    = {36},
  number    = {4},
  pages     = {500--510},
  year      = {2025},
  month     = {July},
  doi       = {10.1097/EDE.0000000000001863},
  url       = {https://www.ovid.com/jnls/epidem/fulltext/10.1097/ede.0000000000001863~generalizing-and-transporting-causal-inferences-from}
}

@article{Dahabreh2019,
  author = {Dahabreh, Issa J. and Robertson, Sarah E. and Tchetgen, Eric J. and Stuart, Elizabeth A. and Hern{\'a}n, Miguel A.},
  title = {Generalizing causal inferences from individuals in randomized trials to all trial-eligible individuals},
  journal = {Biometrics},
  year = {2019},
  volume = {75},
  number = {2},
  pages = {685--694},
  doi = {10.1111/biom.13009},
  url = {https://doi.org/10.1111/biom.13009}
}

@article{dorie2016flexible,
  author = {Dorie, Vincent and Harada, Masataka and Carnegie, Nicole Bohme and Hill, Jennifer},
  title = {A Flexible, Interpretable Framework for Assessing Sensitivity to Unmeasured Confounding},
  journal = {Statistics in Medicine},
  volume = {35},
  number = {20},
  pages = {3453--3470},
  year = {2016},
  doi = {10.1002/sim.6973},
  url = {https://doi.org/10.1002/sim.6973}
}

@inproceedings{titsias2009variational,
  title     = {Variational Learning of Inducing Variables in Sparse Gaussian Processes},
  author    = {Titsias, Michalis},
  booktitle = {Proceedings of the Twelfth International Conference on Artificial Intelligence and Statistics},
  volume    = {5},
  pages     = {567--574},
  year      = {2009},
  publisher = {PMLR},
  url       = {https://proceedings.mlr.press/v5/titsias09a.html}
}

\newpage
\appendix
\section{Integration over a Target Covariate Distribution}
\label{app:ate}

Let \(P\) denote a target covariate distribution on \(\mathcal X\). Under Assumption~\ref{A5:S_exchangeability}, \(\nu_{\mathrm{e}}(x,a)=\nu_{\mathrm{o}}(x,a)\), hence
\[
\mathbb E_{X \sim P} [\nu_{\mathrm{e}}(X,a)] = \mathbb E_{X \sim P} [\nu_{\mathrm{o}}(X,a)].
\]

The target-population estimate of \(\mu_{\mathrm{e}}\) under \emph{Causal-ICM} is \(\int_\mathcal{X} m_{\mathrm{e}|\mathcal{D}}(x)\, dP(x)\). Using the pointwise decomposition \(m_{\mathrm{e}|\mathcal{D}}(x) = m_{\mathrm{e}|\mathcal{D}_{\mathrm{e}}}(x) + C(x;\rho)\),
\[
\int_\mathcal{X} m_{\mathrm{e}|\mathcal{D}}(x)\, dP(x) = \int_\mathcal{X} m_{\mathrm{e}|\mathcal{D}_{\mathrm{e}}}(x)\, dP(x) + \int_\mathcal{X} C(x;\rho)\, dP(x).
\]
Similarly,
\[
\int_\mathcal{X} V_{\mathrm{e}|\mathcal{D}}(x)\, dP(x) = \int_\mathcal{X} V_{\mathrm{e}|\mathcal{D}_{\mathrm{e}}}(x)\, dP(x) - \int_\mathcal{X} R(x;\rho)\, dP(x).
\]

\begin{proposition}
\label{prop:v_bound_ate}
For \(\rho \in (0,1)\),

\[
\int_\mathcal{X} V_{\mathrm{e}|\mathcal{D}}(x)\, dP(x) \geq (1-\rho^2) \int_\mathcal{X} V_{\mathrm{e}|\mathcal{D}_{\mathrm{e}}}(x)\, dP(x).
\]
\end{proposition}

\begin{proof}
Proposition~\ref{prop:v_bound} gives \(V_{\mathrm{e}|\mathcal D}(x)\ge(1-\rho^2)V_{\mathrm{e}|\mathcal D_{\mathrm{e}}}(x)\) for every \(x\in\mathcal X\). Integrating both sides with respect to \(P\) and using monotonicity and linearity of the integral gives the result.
\end{proof}

Since \(\int_\mathcal{X} R(x;\rho)\, dP(x)\geq 0\),
\[
\int_\mathcal{X} V_{\mathrm{e}|\mathcal{D}}(x)\, dP(x) \leq \int_\mathcal{X} V_{\mathrm{e}|\mathcal{D}_{\mathrm{e}}}(x)\, dP(x).
\]
\section{Tipping-Point Analysis}
\label{app:tipping-point}

Fix \(\rho\) and treatment arm \(a\). Let \(\mu_{\mathrm{e}}(x;\rho,a)\mid\mathcal D\) denote the posterior experimental response surface under arm \(a\), with mean and covariance
\[
m_{\mathrm{e}|\mathcal D}(x;\rho,a) = \mathbb E[\mu_{\mathrm{e}}(x;\rho,a)\mid\mathcal D], \qquad K_{\mathrm{e}|\mathcal D}(x,x';\rho,a) = \operatorname{Cov}\{\mu_{\mathrm{e}}(x;\rho,a), \mu_{\mathrm{e}}(x';\rho,a)\mid\mathcal D\}.
\]
For a target covariate distribution \(P\), define
\[
\psi(\rho;a) = \int_\mathcal X \mu_{\mathrm{e}}(x;\rho,a)\, dP(x).
\]
As a linear functional of a \gls{gp}, \(\psi(\rho;a)\mid\mathcal D \sim \mathcal N\big(\mu_\psi(\rho;a), \sigma_\psi^2(\rho;a)\big)\), with
\[
\mu_\psi(\rho;a) = \int_\mathcal X m_{\mathrm{e}|\mathcal D}(x;\rho,a)\, dP(x), \qquad \sigma_\psi^2(\rho;a) = \iint_{\mathcal X^2} K_{\mathrm{e}|\mathcal D}(x,x';\rho,a)\, dP(x)\, dP(x').
\]
Under independently modelled arm-specific surfaces, \(\Psi(\rho) = \psi(\rho;1)-\psi(\rho;0)\) and
\[
\Psi(\rho)\mid\mathcal D \sim \mathcal N\big(\mu_\Psi(\rho), \sigma_\Psi^2(\rho)\big), \qquad \mu_\Psi(\rho)=\mu_\psi(\rho;1)-\mu_\psi(\rho;0), \qquad \sigma_\Psi^2(\rho)=\sigma_\psi^2(\rho;1)+\sigma_\psi^2(\rho;0).
\]

For a two-sided \(100(1-\alpha)\%\) credible interval \(\text{CI}_{1-\alpha}(\rho)=[L_\alpha(\rho),U_\alpha(\rho)]\),
\[
L_\alpha(\rho) = \mu_\Psi(\rho) - z_{1-\alpha/2}\sigma_\Psi(\rho), \qquad U_\alpha(\rho) = \mu_\Psi(\rho) + z_{1-\alpha/2}\sigma_\Psi(\rho).
\]
We assess whether the posterior credible interval excludes zero. An interval entirely above zero supports a positive effect, while an interval entirely below zero supports a negative effect. We report the ranges of $\rho$ over which each conclusion holds.

A tipping point is a value of $\rho$ at which an endpoint of the
credible interval meets zero:
\[
\mathcal{T}
=
\{\rho \in [0,1] : L_\alpha(\rho)=0
\ \text{or}\ U_\alpha(\rho)=0\}.
\]
If there is one tipping point, we denote it by $\rho^*$; if there
are several, we report each and describe the conclusion on either
side. At a tipping point, the credible interval includes zero.
If the interval excludes zero only for $\rho>\rho^*$, the tipping
point marks the degree of borrowing from the observational data
above which the posterior supports a nonzero effect. If it excludes
zero only for $\rho<\rho^*$, increasing borrowing beyond the tipping
point removes that support. For a negative effect, the relevant
endpoint is $U_\alpha$; for a positive effect, it is $L_\alpha$.



Candidate tipping points can be located by evaluating \(\mu_\Psi(\rho)\) and \(\sigma_\Psi(\rho)\) on a grid over \([0,1]\), locating sign changes in \(L_\alpha\) or \(U_\alpha\), and refining by bisection. \(L_\alpha(\rho)\) and \(U_\alpha(\rho)\) are continuous in \(\rho\) wherever the relevant covariance matrices are positive definite.

For \(P\) represented empirically by points \(\tilde x_1,\dots,\tilde x_M\) with weights \(w_1,\dots,w_M\) such that \(\sum_{i=1}^M w_i=1\),
\[
\mu_\psi(\rho;a) = \sum_{i=1}^M w_i\, m_{\mathrm{e}|\mathcal D}(\tilde x_i;\rho,a), \qquad \sigma_\psi^2(\rho;a) = \sum_{i=1}^M\sum_{j=1}^M w_i w_j\, K_{\mathrm{e}|\mathcal D}(\tilde x_i,\tilde x_j;\rho,a),
\]
or, with \(w=(w_1,\dots,w_M)^\top\), \(m_{\mathrm{e}|\mathcal D}\) the vector of posterior means and \(K_{\mathrm{e}|\mathcal D}\) the posterior covariance matrix at the target points,
\[
\mu_\psi(\rho;a) = w^\top m_{\mathrm{e}|\mathcal D}, \qquad \sigma_\psi^2(\rho;a) = w^\top K_{\mathrm{e}|\mathcal D}\, w.
\]

\section{Proofs}

\subsection{Proof of Proposition \ref{prop:joint_post_distr}}\label{app:proof:joint_post_distr}

\begin{proof}
Write \(m_{\mathrm{e}|\mathcal D_{\mathrm{e}}}(x_*)=k(x_*,X_{\mathrm{e}})\Sigma_{\mathrm{e}}^{-1}y_{\mathrm{e}}\) and \(V_{\mathrm{e}|\mathcal D_{\mathrm{e}}}(x_*)=k(x_*,x_*)-k'(x_*,x_*)\) for the posterior mean and variance of \(\mu_{\mathrm{e}}(x_*)\) given the trial data alone, and suppress the argument \(x_*\) in \(q\), \(a_{\mathrm{o}}\) and these quantities below.

Condition on \(y_{\mathrm{e}}\) first. The vector \((\mu_{\mathrm{e}}(x_*),\mu_{\mathrm{o}}(x_*),y_{\mathrm{o}})\) is jointly Gaussian with \(y_{\mathrm{e}}\), because it is a linear functional of the Gaussian processes plus independent Gaussian noise. The prior covariances follow from \(\operatorname{Cov}(\mu_s(x),\mu_{s'}(x'))=B_{ss'}k(x,x')\) with \(B=\begin{pmatrix}1&\rho\\ \rho&1\end{pmatrix}\). In particular,
\begin{gather*}
\operatorname{Cov}(\mu_{\mathrm{e}}(x_*),y_{\mathrm{e}})=k(x_*,X_{\mathrm{e}}),\qquad \operatorname{Cov}(\mu_{\mathrm{o}}(x_*),y_{\mathrm{e}})=\rho\,k(x_*,X_{\mathrm{e}}),\qquad \operatorname{Cov}(y_{\mathrm{o}},y_{\mathrm{e}})=\rho\,k(X_{\mathrm{o}},X_{\mathrm{e}}),\\
\operatorname{Cov}(\mu_{\mathrm{e}}(x_*),y_{\mathrm{o}})=\rho\,k(x_*,X_{\mathrm{o}}),\qquad \operatorname{Cov}(\mu_{\mathrm{o}}(x_*),y_{\mathrm{o}})=k(x_*,X_{\mathrm{o}}),\qquad \operatorname{Cov}(\mu_{\mathrm{e}}(x_*),\mu_{\mathrm{o}}(x_*))=\rho\,k(x_*,x_*).
\end{gather*}
Gaussian conditioning on \(y_{\mathrm{e}}\) replaces each mean by \(\operatorname{Cov}(\cdot,y_{\mathrm{e}})\Sigma_{\mathrm{e}}^{-1}y_{\mathrm{e}}\) and each covariance \(\operatorname{Cov}(a,b)\) by \(\operatorname{Cov}(a,b)-\operatorname{Cov}(a,y_{\mathrm{e}})\Sigma_{\mathrm{e}}^{-1}\operatorname{Cov}(y_{\mathrm{e}},b)\). The conditional means are therefore
\[
\mathbb E[\mu_{\mathrm{e}}(x_*)\mid y_{\mathrm{e}}]=m_{\mathrm{e}|\mathcal D_{\mathrm{e}}},\qquad
\mathbb E[\mu_{\mathrm{o}}(x_*)\mid y_{\mathrm{e}}]=\rho\,m_{\mathrm{e}|\mathcal D_{\mathrm{e}}},\qquad
\mathbb E[y_{\mathrm{o}}\mid y_{\mathrm{e}}]=\rho\,k(X_{\mathrm{o}},X_{\mathrm{e}})\Sigma_{\mathrm{e}}^{-1}y_{\mathrm{e}},
\]
and the conditional covariances are
\begin{align*}
\operatorname{Var}(\mu_{\mathrm{e}}(x_*)\mid y_{\mathrm{e}})&=V_{\mathrm{e}|\mathcal D_{\mathrm{e}}}, \\
\operatorname{Cov}(\mu_{\mathrm{e}}(x_*),\mu_{\mathrm{o}}(x_*)\mid y_{\mathrm{e}})&=\rho\,V_{\mathrm{e}|\mathcal D_{\mathrm{e}}}, \\
\operatorname{Var}(\mu_{\mathrm{o}}(x_*)\mid y_{\mathrm{e}})&=k(x_*,x_*)-\rho^2k'(x_*,x_*), \\
\operatorname{Cov}(\mu_{\mathrm{e}}(x_*),y_{\mathrm{o}}\mid y_{\mathrm{e}})&=\rho\,k(x_*,X_{\mathrm{o}})-\rho\,k'(x_*,X_{\mathrm{o}})=\rho\,q, \\
\operatorname{Cov}(\mu_{\mathrm{o}}(x_*),y_{\mathrm{o}}\mid y_{\mathrm{e}})&=k(x_*,X_{\mathrm{o}})-\rho^2k'(x_*,X_{\mathrm{o}})=a_{\mathrm{o}}, \\
\operatorname{Var}(y_{\mathrm{o}}\mid y_{\mathrm{e}})&=\Sigma_{\mathrm{eo}}.
\end{align*}
Here \(\Sigma_{\mathrm{eo}}\succeq\sigma^2I\) is invertible, and \(y_{\mathrm{o}}-\mathbb E[y_{\mathrm{o}}\mid y_{\mathrm{e}}]=r_{\mathrm{o}}\).

Since \(p(\boldsymbol{\mu}(x_*)\mid y_{\mathrm{e}},y_{\mathrm{o}})\propto p(\boldsymbol{\mu}(x_*),y_{\mathrm{o}}\mid y_{\mathrm{e}})\), the posterior given \(\mathcal D\) is obtained by conditioning on \(y_{\mathrm{o}}\) under the Gaussian law above. It is therefore bivariate normal, with mean
\[
\begin{pmatrix}m_{\mathrm{e}|\mathcal D_{\mathrm{e}}}\\ \rho\,m_{\mathrm{e}|\mathcal D_{\mathrm{e}}}\end{pmatrix}+\begin{pmatrix}\rho\,q\\ a_{\mathrm{o}}\end{pmatrix}\Sigma_{\mathrm{eo}}^{-1}r_{\mathrm{o}}
\]
and covariance
\[
\begin{pmatrix}V_{\mathrm{e}|\mathcal D_{\mathrm{e}}}&\rho\,V_{\mathrm{e}|\mathcal D_{\mathrm{e}}}\\ \rho\,V_{\mathrm{e}|\mathcal D_{\mathrm{e}}}&k(x_*,x_*)-\rho^2k'(x_*,x_*)\end{pmatrix}-\begin{pmatrix}\rho\,q\\ a_{\mathrm{o}}\end{pmatrix}\Sigma_{\mathrm{eo}}^{-1}\begin{pmatrix}\rho\,q\\ a_{\mathrm{o}}\end{pmatrix}^{\!\top}.
\]
Reading off the entries gives

\begin{align*}
m_{\mathrm{e}|\mathcal D}(x_*)&=m_{\mathrm{e}|\mathcal D_{\mathrm{e}}}+\rho\,q\,\Sigma_{\mathrm{eo}}^{-1}r_{\mathrm{o}}, &
V_{\mathrm{e}|\mathcal D}(x_*)&=V_{\mathrm{e}|\mathcal D_{\mathrm{e}}}-\rho^2\,q\,\Sigma_{\mathrm{eo}}^{-1}q^\top,\\
m_{\mathrm{o}|\mathcal D}(x_*)&=\rho\,m_{\mathrm{e}|\mathcal D_{\mathrm{e}}}+a_{\mathrm{o}}\,\Sigma_{\mathrm{eo}}^{-1}r_{\mathrm{o}}, &
V_{\mathrm{o}|\mathcal D}(x_*)&=k(x_*,x_*)-\rho^2k'(x_*,x_*)-a_{\mathrm{o}}\,\Sigma_{\mathrm{eo}}^{-1}a_{\mathrm{o}}^\top,
\end{align*}
\[
C_{\mathrm{eo}|\mathcal D}(x_*)=\rho\,V_{\mathrm{e}|\mathcal D_{\mathrm{e}}}-\rho\,q\,\Sigma_{\mathrm{eo}}^{-1}a_{\mathrm{o}}^\top,
\]
as stated.
\end{proof}


\subsection{Proof of Corollary \ref{cor:post_icm_obs_only}}\label{app:proof:post_icm_obs_only}

\begin{proof}
Set \(\mathcal D_{\mathrm{e}}=\emptyset\) in Proposition~\ref{prop:joint_post_distr}. Then \(k'\equiv0\), \(q(x_*)=a_{\mathrm{o}}(x_*)=k(x_*,X_{\mathrm{o}})\), \(\Sigma_{\mathrm{eo}}=\Sigma_{\mathrm{o}}\) and \(r_{\mathrm{o}}=y_{\mathrm{o}}\). Hence
\(m_{\mathrm{o}}(x_*)=k(x_*,X_{\mathrm{o}})\Sigma_{\mathrm{o}}^{-1}y_{\mathrm{o}}\) and \(V_{\mathrm{o}}(x_*)=k(x_*,x_*)-k(x_*,X_{\mathrm{o}})\Sigma_{\mathrm{o}}^{-1}k(X_{\mathrm{o}},x_*)\),
while \(\mathbb E[\mu_{\mathrm{e}}(x_*)\mid\mathcal D_{\mathrm{o}}]=\rho\,m_{\mathrm{o}}(x_*)\) and
\(\operatorname{Var}(\mu_{\mathrm{e}}(x_*)\mid\mathcal D_{\mathrm{o}})=k(x_*,x_*)-\rho^2\{k(x_*,x_*)-V_{\mathrm{o}}(x_*)\}=(1-\rho^2)k(x_*,x_*)+\rho^2V_{\mathrm{o}}(x_*)\),
using \(k(x_*,X_{\mathrm{o}})\Sigma_{\mathrm{o}}^{-1}k(X_{\mathrm{o}},x_*)=k(x_*,x_*)-V_{\mathrm{o}}(x_*)\).
\end{proof}

\begin{proof}[Direct verification]
The relevant prior covariances are
\[
\operatorname{Cov}\!\left(\mu_{\mathrm{o}}(x_*),y_{\mathrm{o}}\right)=k(x_*,X_{\mathrm{o}}),
\qquad
\operatorname{Cov}\!\left(\mu_{\mathrm{e}}(x_*),y_{\mathrm{o}}\right)=\rho\,k(x_*,X_{\mathrm{o}}),
\]
the second following from \(\operatorname{Cov}(u_1(x_*),\mu_{\mathrm{o}}(X_{\mathrm{o}}))=\rho\,k(x_*,X_{\mathrm{o}})\).
Both \(m_{\mathrm{o}}(x_*),V_{\mathrm{o}}(x_*)\) are the standard Gaussian-conditioning posterior of \(\mu_{\mathrm{o}}(x_*)\) given
\(y_{\mathrm{o}}\sim\mathcal N(0,\Sigma_{\mathrm{o}})\). For \(\mu_{\mathrm{e}}(x_*)\), joint Gaussianity and
\(\operatorname{Cov}(\mu_{\mathrm{e}}(x_*),y_{\mathrm{o}})=\rho\,k(x_*,X_{\mathrm{o}})\) give
\[
\mathbb E[\mu_{\mathrm{e}}(x_*)\mid\mathcal D_{\mathrm{o}}]=\rho\,k(x_*,X_{\mathrm{o}})\Sigma_{\mathrm{o}}^{-1}y_{\mathrm{o}}=\rho\,m_{\mathrm{o}}(x_*),
\]
\[
\operatorname{Var}(\mu_{\mathrm{e}}(x_*)\mid\mathcal D_{\mathrm{o}})
=k(x_*,x_*)-\rho^2 k(x_*,X_{\mathrm{o}})\Sigma_{\mathrm{o}}^{-1}k(X_{\mathrm{o}},x_*)
=(1-\rho^2)k(x_*,x_*)+\rho^2 V_{\mathrm{o}}(x_*),
\]
using \(k(x_*,X_{\mathrm{o}})\Sigma_{\mathrm{o}}^{-1}k(X_{\mathrm{o}},x_*)=k(x_*,x_*)-V_{\mathrm{o}}(x_*)\). \qedhere
\end{proof}


\subsection{Proof of Proposition \ref{prop:obs_shift_bounds}} \label{app: proof: obs_shift_bounds}

\begin{proof}
Write \(\mathbb E^\star\) and \(\operatorname{Cov}^\star\) for expectation and covariance under \(P_{\mathrm{o}}^\star\).
Write \(C=C(x_*;\rho)\), \(R=R(x_*;\rho)\), \(q=q(x_*)\) and
\(V=V_{\mathrm{e}|\mathcal D_{\mathrm{e}}}(x_*)\). Let \(K_{st}=k(X_s,X_t)\) for
\(s,t\in\{\mathrm{e},\mathrm{o}\}\). The trial-only posterior mean function is
\(m_{\mathrm{e}|\mathcal D_{\mathrm{e}}}(x)=k(x,X_{\mathrm{e}})\Sigma_{\mathrm{e}}^{-1}y_{\mathrm{e}}\).
Under the working model, conditioning on the trial data and both designs
gives the predictive mean and covariance of \(y_{\mathrm{o}}\) as
\[
  m_{\mathrm{o}|\mathcal D_{\mathrm{e}}}(X_{\mathrm{o}})=\rho K_{\mathrm{oe}}\Sigma_{\mathrm{e}}^{-1}y_{\mathrm{e}},
  \qquad
  \Sigma_{\mathrm{eo}}=K_{\mathrm{oo}}+\sigma^2 I
    -\rho^2K_{\mathrm{oe}}\Sigma_{\mathrm{e}}^{-1}K_{\mathrm{eo}}.
\]
The conditional cross-covariance of \(\mu_{\mathrm{e}}(x_*)\) with \(y_{\mathrm{o}}\) is
\(\rho q\), where
\(q=k(x_*,X_{\mathrm{o}})-k(x_*,X_{\mathrm{e}})\Sigma_{\mathrm{e}}^{-1}K_{\mathrm{eo}}\).
With \(r_{\mathrm{o}}=y_{\mathrm{o}}-m_{\mathrm{o}|\mathcal D_{\mathrm{e}}}(X_{\mathrm{o}})\), Gaussian conditioning on
\(y_{\mathrm{o}}\) gives
\[
  C=\rho q\Sigma_{\mathrm{eo}}^{-1}r_{\mathrm{o}},
  \qquad R=\rho^2q\Sigma_{\mathrm{eo}}^{-1}q^\top.
\]

\noindent\textbf{Step 1. Bound the variance reduction.}
Define
\[
  T=K_{\mathrm{oe}}\Sigma_{\mathrm{e}}^{-1}K_{\mathrm{eo}},
  \qquad S_0=K_{\mathrm{oo}}-T+\sigma^2 I.
\]
We have \(T\succeq0\). The matrix \(K_{\mathrm{oo}}-T\) is the Schur complement
of \(\Sigma_{\mathrm{e}}\) in the joint covariance of noisy trial evaluations
and noiseless observational evaluations of a single \gls{gp} with kernel
\(k\), so it is positive semidefinite. Consequently
\[
  \Sigma_{\mathrm{eo}}=S_0+(1-\rho^2)T\succeq S_0\succeq \sigma^2 I,
  \qquad \Sigma_{\mathrm{eo}}^{-1}\preceq S_0^{-1}.
\]
At \(\rho=1\), the conditional covariance of
\((\mu_{\mathrm{e}}(x_*),y_{\mathrm{o}})\) given \(X_{\mathrm{o}},\mathcal D_{\mathrm{e}}\) has blocks
\(\bigl(\begin{smallmatrix}V&q\\ q^\top&S_0\end{smallmatrix}\bigr)\).
The Schur complement of \(S_0\) therefore gives
\(qS_0^{-1}q^\top\le V\). It follows that
\begin{equation}
\label{eq:observational-variance-reduction}
  0\le R=\rho^2q\Sigma_{\mathrm{eo}}^{-1}q^\top
    \le \rho^2qS_0^{-1}q^\top\le \rho^2 V.
\end{equation}
For any vector \(v\in\mathbb R^{n_{\mathrm{o}}}\), Cauchy--Schwarz gives
\[
  |\rho q\Sigma_{\mathrm{eo}}^{-1}v|
    \le \sqrt R\,(v^\top\Sigma_{\mathrm{eo}}^{-1}v)^{1/2}.
\]
Taking \(v=r_{\mathrm{o}}\) gives \(|C|\le\sqrt R\,(r_{\mathrm{o}}^\top\Sigma_{\mathrm{eo}}^{-1}r_{\mathrm{o}})^{1/2}\).
Since \(\Sigma_{\mathrm{eo}}\succ0\), \(R=0\) implies \(\rho q=0\) and hence
\(C=0\).

\noindent\textbf{Step 2. Split the mean shift and bound the bias.}
Under (B1), define
\(h=\mu_{\mathrm{o}}-\rho m_{\mathrm{e}|\mathcal D_{\mathrm{e}}}\). The trial-only posterior mean is a
finite linear combination of kernel sections at \(X_{\mathrm{e}}\), so
\(h\in\mathcal H_k\). Hence
\[
  B_h^2=\|h\|_{\mathcal H_k}^2+\sigma^{-2}\|h(X_{\mathrm{e}})\|_2^2
\]
is finite and, for the fixed trial, independent of \(X_{\mathrm{o}}\) and \(n_{\mathrm{o}}\).
Conditional on \(X_{\mathrm{o}},\mathcal D_{\mathrm{e}}\), set \(b=h(X_{\mathrm{o}})\) and
\(w=\rho q\Sigma_{\mathrm{eo}}^{-1}\). Since \(r_{\mathrm{o}}=b+\varepsilon_{\mathrm{o}}\), the shift
splits as \(C=wb+w\varepsilon_{\mathrm{o}}\). To bound the contribution of \(b\),
stack the inputs as \(X_{\mathrm{all}}=(X_{\mathrm{e}},X_{\mathrm{o}})\), write
\(K=k(X_{\mathrm{all}},X_{\mathrm{all}})\), and set
\[
  v=\begin{pmatrix}0\\ b\end{pmatrix}=h(X_{\mathrm{all}})+d,
  \qquad d=\begin{pmatrix}-h(X_{\mathrm{e}})\\0\end{pmatrix}.
\]
Let \(\gamma=(K+\sigma^2 I)^{-1}v\), \(Q=v^\top\gamma\), and
\(g=\sum_i\gamma_i k(\cdot,x_i)\), where \(x_i\) ranges over the
stacked inputs. The reproducing property gives
\[
  Q=\|g\|_{\mathcal H_k}^2+\sigma^2\|\gamma\|_2^2,
  \qquad
  Q=\langle h,g\rangle_{\mathcal H_k}+d^\top\gamma.
\]
Cauchy--Schwarz applied to the two terms of the second expression gives
\(Q\le (\|h\|_{\mathcal H_k}^2+\sigma^{-2}\|d\|_2^2)^{1/2}\sqrt Q\),
and hence \(Q\le B_h^2\), including when \(Q=0\). The lower-right
block of \((K+\sigma^2 I)^{-1}\) is \(S_0^{-1}\) by the block inverse
formula. Thus
\[
  b^\top\Sigma_{\mathrm{eo}}^{-1}b\le b^\top S_0^{-1}b=Q\le B_h^2.
\]
The Cauchy--Schwarz bound from Step~1 then gives \(|wb|\le\sqrt R\,B_h\).

\noindent\textbf{Step 3. Bound the noise contribution.}
The regression definition of \(\mu_{\mathrm{o}}\) gives mean-zero residuals. By
(B2), the residuals remain i.i.d.\ with their common law
conditional on \(X_{\mathrm{o}},\mathcal D_{\mathrm{e}}\). Thus
\[
  \mathbb E^\star[\varepsilon_{\mathrm{o}}\mid X_{\mathrm{o}},\mathcal D_{\mathrm{e}}]=0,
  \qquad
  \operatorname{Cov}^\star(\varepsilon_{\mathrm{o}}\mid X_{\mathrm{o}},\mathcal D_{\mathrm{e}})
    =\sigma_\varepsilon^2 I.
\]
Since \(\Sigma_{\mathrm{eo}}\succeq\sigma^2 I\) by Step~1, the conditional
variance of \(w\varepsilon_{\mathrm{o}}\) satisfies
\[
  \sigma_\varepsilon^2 ww^\top
    \le \frac{\sigma_\varepsilon^2}{\sigma^2}w\Sigma_{\mathrm{eo}}w^\top
    = \frac{\sigma_\varepsilon^2}{\sigma^2}R.
\]

\noindent\textbf{Step 4. Combine the bounds.}
Conditional centring makes the cross term in \(C^2\) vanish. Combining
the bounds on the bias and noise terms with
\eqref{eq:observational-variance-reduction} gives the finite-sample bound
\[
  \bigl\{\mathbb E^\star[C^2\mid X_{\mathrm{o}},\mathcal D_{\mathrm{e}}]\bigr\}^{1/2}
  \le \sqrt{R}\,\bigl(B_h^2+\sigma_\varepsilon^2/\sigma^2\bigr)^{1/2}
  \le \rho\sqrt V\,\bigl(B_h^2+\sigma_\varepsilon^2/\sigma^2\bigr)^{1/2}.
\]
When \(\rho\sqrt V>0\), Markov's inequality gives, for every \(t>0\),
\[
  P_{\mathrm{o}}^\star\bigl(|C|>t\rho\sqrt V\mid X_{\mathrm{o}},\mathcal D_{\mathrm{e}}\bigr)
    \le \frac{B_h^2+\sigma_\varepsilon^2/\sigma^2}{t^2}.
\]
The right-hand side is independent of \(n_{\mathrm{o}}\) and \(X_{\mathrm{o}}\), proving the
uniform stochastic bound. If \(\rho\sqrt V=0\), Step~1 gives \(R=0\) and
hence \(C=0\).
\end{proof}

\paragraph{Interpretation.}
With the trial data and working model parameters held fixed, the
root-mean-square change in the posterior mean at \(x_*\) is bounded
uniformly as the observational sample grows. The scale of the bound is
\(\rho\) times the trial-only posterior standard deviation. Its
multiplier accounts for both the discrepancy between the true
observational regression and the trial-based prediction, through
\(B_h\), and the observational error variance, through
\(\sigma_\varepsilon^2/\sigma^2\). Large discrepancies or noise can therefore still allow
a substantial shift. At \(\rho=0\), the observational data leave the
posterior mean unchanged.

The i.i.d.\ errors describe individual outcome variation around \(\mu_{\mathrm{o}}\).
Unmeasured confounding can affect \(\mu_{\mathrm{o}}\) and hence the bias term even
with independent errors. With a random design, the same probability
bound applies after averaging over the observational covariates while
holding the trial data fixed.

\paragraph{Remark (Gaussian predictive law).}
If \(y_{\mathrm{o}}\mid X_{\mathrm{o}},\mathcal D_{\mathrm{e}}\) follows the \emph{Causal-ICM} predictive law
at the analysis value of \(\rho\), then
\[
  C(x_*;\rho)\mid X_{\mathrm{o}},\mathcal D_{\mathrm{e}}
    \sim \mathcal N\bigl(0,R(x_*;\rho)\bigr),
\]
with a point mass at zero when \(R(x_*;\rho)=0\). Indeed, under this law
\(r_{\mathrm{o}}\mid X_{\mathrm{o}},\mathcal D_{\mathrm{e}}\sim\mathcal N(0,\Sigma_{\mathrm{eo}})\), and the
linear map \(r_{\mathrm{o}}\mapsto\rho q\Sigma_{\mathrm{eo}}^{-1}r_{\mathrm{o}}\) has conditional
variance \(R(x_*;\rho)\). Hence for \(0<\alpha<1\),
\[
  \Pr\!\left(
    |C(x_*;\rho)|\le z_{1-\alpha/2}\,
      \rho\sqrt{V_{\mathrm{e}|\mathcal D_{\mathrm{e}}}(x_*)}
    \,\middle|\, X_{\mathrm{o}},\mathcal D_{\mathrm{e}}
  \right)\ge 1-\alpha,
\]
where \(z_{1-\alpha/2}\) is the standard normal quantile.

\paragraph{Interpretation.}
Before seeing the observational outcomes, hold the trial data and
observational covariates fixed. Under the working model at the chosen
\(\rho\), the updated mean averages to the trial-only mean across
possible observational outcomes. The amount of posterior variance
removed, \(R\), is exactly the predictive variance of the posterior mean shift.
Thus a design that yields a larger reduction in posterior uncertainty
also yields a larger standard deviation for the shift. The probability
bound controls the magnitude of the shift under the model's predictive law.

\subsection{Proof of Proposition \ref{prop:v_bound}} \label{app: proof: v_bound}

\begin{proof}
By the posterior variance decomposition and
\eqref{eq:observational-variance-reduction} in the proof of Proposition~\ref{prop:obs_shift_bounds},
\[
  V_{\mathrm{e}|\mathcal D}(x_*)
  = V_{\mathrm{e}|\mathcal D_{\mathrm{e}}}(x_*)-R(x_*;\rho)
  \ge (1-\rho^2)V_{\mathrm{e}|\mathcal D_{\mathrm{e}}}(x_*).
\]
\end{proof}

\paragraph{Interpretation.}
Under partial borrowing, observational data can remove at most a
fraction \(\rho^2\) of the trial-only posterior variance at \(x_*\).
For example, at \(\rho=1/2\), at least three quarters of that variance
remains, even as the observational sample grows arbitrarily large.
This lower bound concerns uncertainty about the trial regression under
the working model. The actual reduction can be smaller and depends on
the observational covariate locations and their conditional covariance
with the trial regression at \(x_*\).

\subsection{Proof of Proposition \ref{prop:bias}}\label{app:proof:osm-bias-identification}

\begin{proof}
    In a single observational study according to the OSM, for each treatment arm \(a\in \{0,1\}\) we have the following quantity:
\begin{equation*}
    \begin{aligned}
        \Delta_{\mathrm{OSM}}(a) 
        &= \int \left\{ \mu(x,u,a)-\mu(x,a) \right\} dP(x,u) \\
        &= \int \mu(x,u,a)dP(x,u) - \int \mu(x,a)dP(x,u) \\
        &= \int \mathbb{E}[Y|X=x, U=u, A=a]dP(x,u) \\
        & \qquad - \int \mathbb{E}[Y|X=x, A=a]dP(x,u)
    \end{aligned}
\end{equation*}
Since both are quantities pertaining to the observational study we can write for clarity
\begin{equation*}
    \begin{aligned}
        \Delta_{\mathrm{OSM}}(a) 
        &= \underbrace{\int \mathbb{E}[Y|X=x, U=u, A=a, S=\mathrm{o}]dP(x,u|S=\mathrm{o})}_{(1)} \\
        & \qquad - \underbrace{\int \mathbb{E}[Y|X=x, A=a, S=\mathrm{o}]dP(x,u|S=\mathrm{o})}_{(2)}
    \end{aligned}
\end{equation*}
The second quantity can be written as 
\begin{equation*}
    \begin{aligned}
        (2) &= \int \mathbb{E}[Y|X=x, A=a, S=\mathrm{o}]dP(x,u|S=\mathrm{o}) \\
        &\quad= \int \left\{ \int dP(u|X=x,S=\mathrm{o}) \right\} \mathbb{E}[Y|X=x, A=a, S=\mathrm{o}]dP(x|S=\mathrm{o}) \\
        &\quad= \int \mathbb{E}[Y|X=x, A=a, S=\mathrm{o}]dP(x|S=\mathrm{o}) 
    \end{aligned}
\end{equation*}
which is identifiable from observed data, since positivity in the observational study (Assumption~\ref{A5:OBS_A_positivity}) gives \(\mathbb{P}(A=a\mid X=x,S=\mathrm{o})>0\) on the support of \(X\mid S=\mathrm{o}\).

Now let us focus on the first quantity

\begin{equation*}
    \begin{aligned}
        (1)&=\int \mathbb{E}[Y|X=x, U=u, A=a, S=\mathrm{o}]dP(x,u|S=\mathrm{o})  \\
        &\quad= \int \mathbb{E}[Y(a)|X=x, U=u, S=\mathrm{o}]dP(x,u|S=\mathrm{o})  \\
        &\qquad\text{\small{(\ref{A1:consistency}, \ref{A4:OBS_A_exchangeability} and \ref{A5:OBS_A_positivity})}}\\
        &\quad= \int \left\{ \int \mathbb{E}[Y(a)|X=x, U=u, S=\mathrm{o}]dP(u|X=x,S=\mathrm{o}) \right\} dP(x|S=\mathrm{o})  \\
        &\quad= \int \mathbb{E}[Y(a)|X=x, S=\mathrm{o}]dP(x|S=\mathrm{o})  \\
        &\qquad\text{\small{(iterated expectation)}}
    \end{aligned}
\end{equation*}
which is the mean of \(Y(a)\) in the observational population. Exchangeability over \(S\) then links it to the \gls{rct},
\begin{equation*}
    \begin{aligned}
        (1) &= \int \mathbb{E}[Y(a)|X=x, S=\mathrm{e}]dP(x|S=\mathrm{o})  \\
        &\qquad\text{\small{(\ref{A5:S_exchangeability}: exchangeability over $S$)}}\\
        &\quad= \int \mathbb{E}[Y|X=x,A=a, S=\mathrm{e}]dP(x|S=\mathrm{o}) \\
        &\qquad\text{\small{(\ref{A1:consistency}--\ref{A3:RCT_A_positivity} and \ref{A6:S_positivity})}}
    \end{aligned}
\end{equation*}
Therefore the first quantity, under the assumptions we are making, is identifiable using RCT data. Subtracting (2) from (1) gives
\begin{equation*}
    \begin{aligned}
        \Delta_{\mathrm{OSM}}(a) &= \int \mathbb{E} \left[ Y \mid X=x,A=a,S=\mathrm{e} \right] \,dP(x \mid S=\mathrm{o}) \\
        & \quad - \int \mathbb{E} \left[ Y \mid X=x,A=a,S=\mathrm{o} \right] \,dP(x \mid S=\mathrm{o}),
    \end{aligned}
\end{equation*}
and since \((1)=\int\mu_{\mathrm{e}}(x,a)\,dP(x|S=\mathrm{o})\), also \(\Delta_{\mathrm{OSM}}(a)=\int\{\mu_{\mathrm{e}}(x,a)-\mu_{\mathrm{o}}(x,a)\}\,dP(x|S=\mathrm{o})=-\int d(x,a)\,dP(x|S=\mathrm{o})\). Finally, \(d(x_*,a)=\mu_{\mathrm{o}}(x_*)-\mu_{\mathrm{e}}(x_*)\) is a linear function of the bivariate normal vector in Proposition~\ref{prop:joint_post_distr}, which gives its posterior.
\end{proof}


\section{Observational-Data-Only Methods by Baseline HbA1c Subgroup}
\label{app:pioneer6-rwd-only}
 
\begin{figure}[htbp]
    \setstretch{1}
    \centering
    \includegraphics[width=0.55\textwidth]{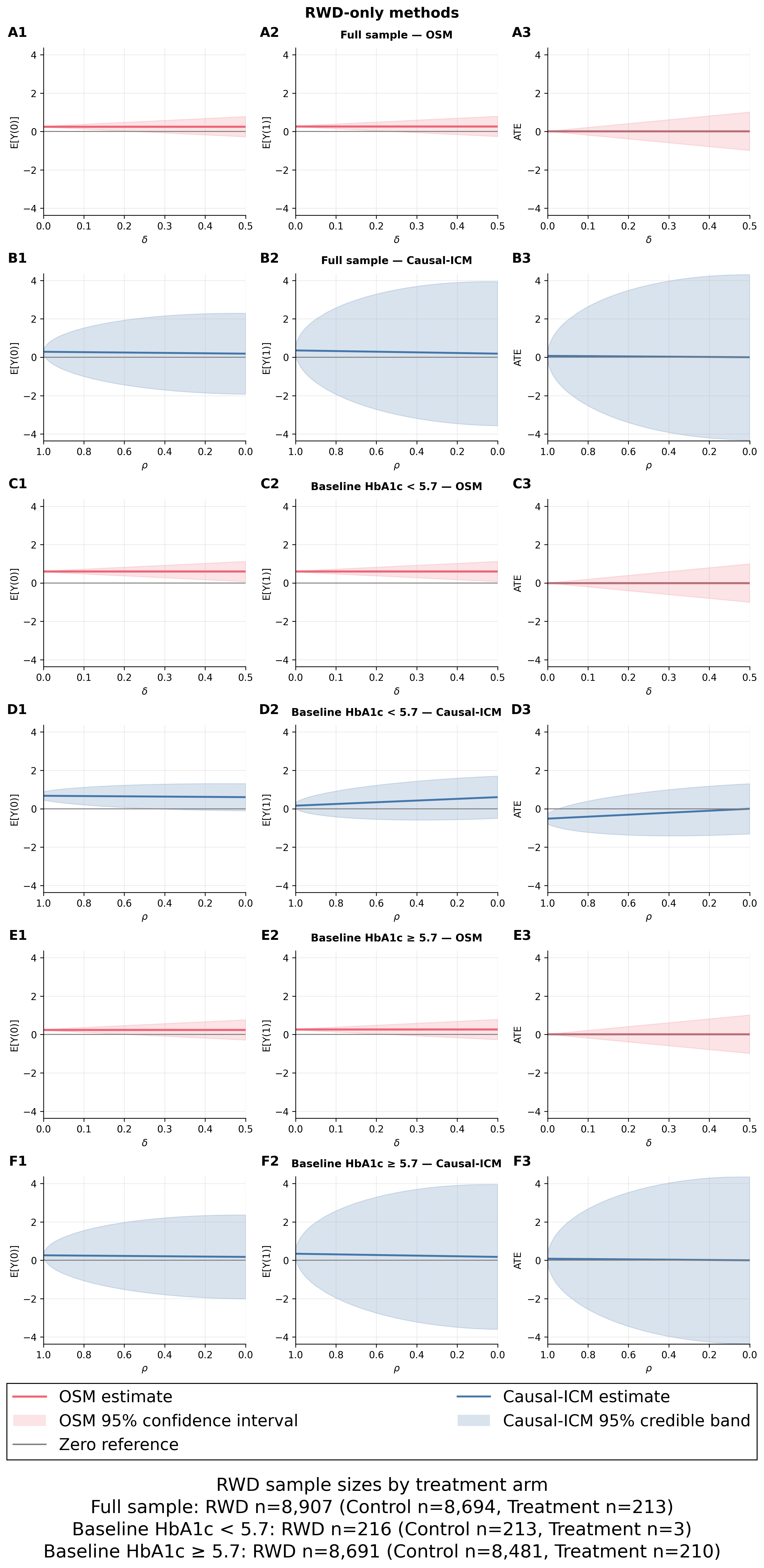}
    \caption{Observational-only sensitivity analyses of full sample and by baseline HbA1c subgroup, for $\mathbb{E}[Y(0)]$, $\mathbb{E}[Y(1)]$ and the ATE (columns 1--3). Rows A, C and E give the OSM, plotted against $\delta$, and rows B, D and F the observational-only \emph{Causal-ICM} posterior, plotted against $\rho$, for the full sample (A, B) and baseline HbA1c $<5.7$ (C, D) and $\geq5.7$ (E, F).}
    \label{fig:pioneer6-rwd-only}
\end{figure}

Web Figure~\ref{fig:pioneer6-rwd-only} reports the OSM and the observational-only \emph{Causal-ICM} posterior (Corollary~\ref{cor:post_icm_obs_only}), for the full sample and separately for the two baseline HbA1c subgroups defined in Section~\ref{sec:pioneer6-application}, using only the claims data. Rows A and B give the OSM and the observational-only \emph{Causal-ICM} posterior for the full sample, rows C and D for patients with baseline HbA1c $<5.7$, and rows E and F for patients with baseline HbA1c $\geq5.7$; columns 1--3 give $\mathbb{E}[Y(0)]$, $\mathbb{E}[Y(1)]$, and the ATE.

The two methods agree for patients with baseline HbA1c $\geq5.7$ but not for those with baseline HbA1c $<5.7$. For patients with baseline HbA1c $<5.7$, the OSM's ATE point estimate is zero at $\delta=0$, giving a tipping point of $\delta^\ast\approx0.00$ (panel C3), whereas the observational-only \emph{Causal-ICM} posterior mean is negative, $\approx-0.52$ at $\rho=1$ with a credible band of $[-0.79,-0.24]$, and shrinks to zero as $\rho$ decreases; the band includes zero for $\rho$ below the tipping point $\rho^\ast\approx0.95$ (panel D3). Only 3 of the 216 claims patients in this subgroup are treated, so the random forest of the OSM cannot split on treatment, and the treated-arm \gls{gp} of \emph{Causal-ICM} rests on three observations and on the covariates that vary among them. For patients with baseline HbA1c $\geq5.7$, the OSM shows the same pattern, with a tipping point of $\delta^\ast\approx0.01$ (panel E3), and the observational-only \emph{Causal-ICM} posterior is markedly less informative: its credible band widens rapidly as $\rho$ decreases, from roughly $[-0.1,0.3]$ at $\rho=1$ to $[-4.4,4.4]$ at $\rho=0$ (panel F3), reflecting greater heterogeneity in the claims population for this subgroup. Without the trial anchor, the claims data therefore give no usable evidence on the effect of semaglutide for patients with baseline HbA1c $\geq5.7$; the sharp conclusions in Section~\ref{sec:pioneer6-application} rely on combining the claims data with the trial.
 
\end{document}